\documentclass[12pt]{article}

\usepackage{braket}
\usepackage{color}
\usepackage{amsmath}
\usepackage{amssymb}
\usepackage{graphicx}
\usepackage{setspace}
\usepackage{slashed}
\usepackage{comment}
\usepackage{cite}
\usepackage{amsthm}
\usepackage{hyperref}
\usepackage{cleveref}
\usepackage[utf8]{inputenc}
\hypersetup{breaklinks}
\usepackage[affil-it]{authblk}

\newtheorem{definition}{Definition}

\makeatletter
\def\@maketitle{%
  \newpage
  \null
  \vskip 2em%
  \begin{center}%
  \let \footnote \thanks
    {\LARGE\bfseries \@title \par}%
    \vskip 2.5em%
    {
      \lineskip .5em%
      \begin{tabular}[t]{c}%
        {\normalsize\@author}
      \end{tabular}\par}%
    \vskip 1.5em%
  \end{center}%
  \par
  \vskip 1.5em}
\makeatother

\newcommand\numberthis{\addtocounter{equation}{1}\tag{\theequation}}

\usepackage[top=1.54cm,bottom=2.54cm,left=2.54cm,right=2.54cm]{geometry}

\numberwithin{equation}{section}

\begin{document}
 \begin{titlepage}

\title{Localized Excitations from Localized Unitary Operators}

\author{Allic Sivaramakrishnan%
  \thanks{\texttt{allic@physics.ucla.edu}}}
\affil{Department of Physics and Astronomy,\\ University of California, Los Angeles,\\ Los Angeles, CA 90095-1547, USA}

\maketitle

\begin{abstract}
Localized unitary operators are basic probes of locality and causality in quantum systems: localized unitary operators create localized excitations in entangled states. Working with an explicit form, we explore properties of these operators in quantum mechanics and quantum field theory. We show that, unlike unitary operators, local non-unitary operators generically create non-local excitations. We present a local picture for quantum systems in which localized experimentalists can only act through localized Hamiltonian deformations, and therefore localized unitary operators. We demonstrate that localized unitary operators model certain quantum quenches exactly. We show how the Reeh-Schlieder theorem follows intuitively from basic properties of entanglement, non-unitary operators, and the local picture. We show that a recent quasi-particle picture for excited-state entanglement entropy in conformal field theories is not universal for all local operators. We prove a causality relation for entanglement entropy and connect our results to the AdS/CFT correspondence.

\end{abstract}

\end{titlepage}
	\tableofcontents

\section{Introduction}

In this work, we study correlation functions in time-dependent states and in theories with time-dependent Hamiltonians. Our results apply to pure states in quantum mechanics and local quantum field theory. Our primary goal is to detail the universal role of localized unitary operators in creating localized excitations. A brief overview of our results is as follows. Just as every observable is represented by some Hermitian operator, every localized excitation is created by some localized unitary operator. If a localized operator is non-unitary, the excitation it creates is not necessarily localized. For example, a localized unitary operator $e^{i\alpha \mathcal{O}(x)}$ creates a localized excitation while the localized non-unitary operator $e^{\alpha \mathcal{O}(x)}$ does not. Localized experimentalists can only act on states by deforming the Hamiltonian by a localized quantity, $H \rightarrow H+H_{loc}(t)$, and this is equivalent to acting with the Heisenberg-picture localized unitary operator $\mathcal{T}\left( e^{i \int dt \mathcal{H}_{loc}(t)} \right)$ on the state. As they create localized excitations, localized unitary operators are tied to basic questions of causality. 

This manuscript extends contemporary studies of excited-state entanglement entropy in conformal field theories (CFTs) \cite{HeNTW14, Nozaki14,Caputa14,NozakiNT14,AsplundBGH115,AsplundBGH215,CaputaV15,CaputaSST14}. Our work builds upon the large body of work on localized excitations and real-time perturbation theory (see, for example, \cite{InInSchwinger, InInKeldysh,ReehSchlieder, KnightLocalization, Licht63, Licht66, Redhead95, Hellwig70, NewtonWigner,Halvorson00, Buchholz15}). In addition to presenting our own results, we reinterpret some well-known results from this body of literature and from quantum information theory that are relevant. We revisit these results to give a coherent picture for the connection between localized unitary operators and localized excitations. We present all results in elementary terms and eschew a complete or axiomatic treatment of the topics discussed. Our purpose is to make contact with modern studies of entanglement entropy in CFT, and a more formal treatment of locality is outside the scope of this work.

While we explore how localized unitary operators create localized excitations, these operators play many other well-known roles in quantum systems. Specific forms of localized unitary operators create squeezed states, coherent states, generalized coherent states, and implement local gauge transformations \cite{Walls83, Ueda93, Gilmore72,Perelomov72,GlauberCoherentState}. Localized unitary operators are also important in large-N and large-dimension limits in quantum mechanics, gauge theories, and the AdS/CFT correspondence.

We now summarize each section. In section 2, we define what we mean by localized and review related concepts. Suppose a Hilbert space $\mathcal{H}$ can be written as a tensor product Hilbert space $\mathcal{H}=\mathcal{H}_1 \otimes \mathcal{H}_2$. An excitation of state $\ket{\Psi} \in \mathcal{H}$ can be represented by acting with some operator $\mathcal{O}_e$ on $\ket{\Psi}$, where $\mathcal{O}_e$ is suitably normalized so that $\mathcal{O}_e\ket{\Psi}$ has unit norm. This excitation is localized in $\mathcal{H}_1$ if 
\begin{equation}
\braket{\Psi|\mathcal{O}^\dagger_e \mathcal{O} \mathcal{O}_e|\Psi} = \braket{\Psi|\mathcal{O}|\Psi}
\end{equation}
for all operators $\mathcal{O}$ local in $\mathcal{H}_2$. $\mathcal{O}$ is local in $\mathcal{H}_2$ when $\mathcal{O}$ can be written as
\begin{equation}
\mathcal{O} \equiv \mathbb{I}_1 \times \mathcal{O}_2.
\end{equation}
Here $\mathbb{I}_1$ is the identity in $\mathcal{H}_1$ and $\mathcal{O}_2: \mathcal{H}_2 \rightarrow \mathcal{H}_2$. In field theory, $\mathcal{H}_1, \mathcal{H}_2$ can be chosen as the Hilbert spaces of the theory restricted to a subregion $A$ of a Cauchy surface and its complement $A^c$. An excitation localized to $A$ does not affect correlation functions of operators inserted at points spacelike-separated from all points in $A$. The familiar local operators $\mathcal{O}(x)$ in field theory are localized to the Hilbert space of every arbitrarily small neighborhood of $x$. Localized operators can be built from operators that are local in different points or Hilbert spaces. For example, if $f(x)$ has support in region $R$, smeared operator $ \int dx f(x) \mathcal{O}(x)$ is localized in $R$. We address subtleties involved in defining localization for gauge theories.

In section 3, we review real-time perturbation theory \cite{InInSchwinger,InInKeldysh}. This perturbation theory gives corrections to correlation functions in time-dependent states perturbatively in a time-dependent interaction Hamiltonian. Perturbation theory for the S-matrix calculates in-out matrix elements $\braket{\Psi_{out}| \Psi_{in}}$, while real-time perturbation theory calculates in-in matrix elements $\braket{\Psi_{in}|\mathcal{O}_1 \ldots \mathcal{O}_n|\Psi_{in}}$. Real-time perturbation theory makes manifest how localized interaction Hamiltonians create localized excitations.

In section 4, we present a coherent picture for time-dependent operations in quantum mechanics and field theory. We call this picture the local picture of quantum systems. In the local picture, an experimentalist can only alter states through deforming the Hamiltonian. A localized experimentalist can only make localized deformations. The excitations that are natural in the local picture are created by acting with a time-ordered localized unitary operator on a state, for example the Heisenberg-picture operator $\mathcal{T}\left(e^{i\int dt J(t)\mathcal{O}(t)} \right)$. As we show later, the familiar local non-unitary operators of field theory generically create non-localized excitations, so the local picture reveals that local experimentalists cannot act with generic local non-unitary operators.

In section 5, we present results in quantum mechanics. We show how local unitary operators alter entangled states locally, while local non-unitary operators alter entangled states non-locally. Local non-unitary operators can be written as state-dependent non-local unitary operators. In the language of quantum information, non-unitary operators implement non-local quantum gates. Our results explain the Reeh-Schlieder theorem intuitively. The superposition of two local excitations may not be a local excitation itself in entangled states. For instance, the sum of two unitary operators $\mathcal{U}_1+\mathcal{U}_2$ is not necessarily unitary. We show that the natural way to combine localized excitations created by operators $\mathcal{U}_1, \mathcal{U}_2$ to produce another localized excitation is by acting with the operators in succession: $\mathcal{U}_1 \mathcal{U}_2$. This prescription for combining localized excitations follows from the local picture.

In section 6, we move on to quantum field theory, our main focus. We show that localized unitary operators create localized excitations. The reason is as follows. If $x,y$ are spacelike-separated, local operators $\mathcal{O}, \mathcal{O}'$ inserted at $x,y$ commute:
\begin{equation}
[\mathcal{O}(x), \mathcal{O}'(y)] = 0.
\end{equation}
It follows immediately that 
\begin{equation}
\braket{\Psi|e^{i \mathcal{O}(x)} \mathcal{O}'(y)  e^{-i\mathcal{O}(x)} |\Psi} = \braket{\Psi|\mathcal{O}'(y) |\Psi}.
\end{equation}
If $x,y$ are not spacelike-separated, then the above equality generically does not hold. For example, in the vacuum of a free real scalar field, the Baker-Campbell-Hausdorff lemma gives
\begin{equation}
\braket{0|e^{i \alpha \phi(x)} \phi(y)  e^{-i\alpha \phi(x)} |0} = i\alpha G_R(x-y)+\ldots,
\end{equation}
where $\alpha$ can be treated as an expansion parameter and $x$ is restricted to the future of $y$. The retarded Green's function $G_R(x-y)$ vanishes when $x-y$ is spacelike. More generally, the commutator of operators $\braket{[\mathcal{O}(x), \mathcal{O}'(y)]}$ diagnoses causality in field theory, and we see that this commutator is in fact the order $\alpha$ correction to $\braket{\mathcal{O}'}$ in the excited state $e^{-i\alpha\mathcal{O}(x)} \ket{\Psi}$. We explore the properties of localized unitary operators, including what we call separable and non-separable localized unitary operators. Separable unitary operators like $e^{i(\mathcal{O}(x)+\mathcal{O}(y))}$ create excitations at $x,y$ that are not entangled with each other. Acting with non-separable unitary operators like $e^{i\mathcal{O}(x) \mathcal{O}(y)}$ create excitations at $x,y$ that may be used to violate causality. As such, non-separable unitary operators cannot be applied to states under time evolution in local quantum field theory. We give a criterion to test separability.

We show that local non-unitary operators can create non-local excitations in field theory. We provide examples and show where the intuition that arbitrary local operators create local excitations breaks down. We provide evidence that certain local non-unitary operators do create local excitations and give examples of others that do not.

In section 7, we apply lessons from the previous sections to give new results concerning the entanglement entropy of excited states in field theory. Recently, a compelling quasi-particle picture has emerged from calculations of entanglement entropy in CFTs \cite{HeNTW14,NozakiNT14,Caputa14,HartmanJK15, AsplundBGH115}. It has been suggested that local operators create entangled pairs of quasi-particles at their insertion point \cite{NozakiNT14}. We provide evidence that this picture applies to local operators with definite conformal dimension. It is known that the quasi-particle picture is invalid for certain theories \cite{Leichenauer15,AsplundBGH115}, and using the example of the operator $e^{\alpha \mathcal{O}(x)}$, we show how this picture fails to extend to all local operators even within theories for which the picture is expected to be accurate. We extend a result in ref. \cite{Headrick14} by proving a general causality relation for entanglement entropy. 

\section{Background: Locality in Quantum Systems}

We will review locality and causality criteria in quantum mechanics and quantum field theory. Causality in field theory is a statement about the commutators of operators. If spacetime points $x,y$ are spacelike-separated, then any two local operators $\mathcal{O}_1(x), \mathcal{O}_2(y)$ commute.
\begin{equation}
[\mathcal{O}_1(x), \mathcal{O}_2(y)] = 0.
\label{eqnCommutator}
\end{equation}
An analogous statement holds if the operators are smeared out over some spacetime region. The vanishing of the commutator for spacelike separation is equivalent to a statement about the branch cuts of all Euclidean correlators that contain $\mathcal{O}_1, \mathcal{O}_2$.

We may state a locality condition based on whether local operations affect observables non-locally. We state a version of this condition first in quantum mechanics. Express the Hilbert space $\mathcal{H}$ of some system as a tensor product Hilbert space $\mathcal{H} = \mathcal{H}_A \otimes \mathcal{H}_B$ of subsystems $A, B$. First we define local operators. 
\begin{definition}
The operator $\mathcal{O}(B)$ is local in $\mathcal{H}_B$ when it can be written as 
\begin{equation}
\mathcal{O}(B) \equiv \mathbb{I}_A \otimes \mathcal{O}_B.
\end{equation} 
Here, $\mathcal{O}_B: \mathcal{H}_B \rightarrow \mathcal{H}_B$ and $\mathbb{I}_A$ is the identity in $\mathcal{H}_A$.
\label{def:def1}
\end{definition}
As we will show in section 5, acting with operator $\mathcal{O}(B)$ may change the expectation value of some operator $\mathcal{O}(A)$ local in $\mathcal{H}_A$, and therefore some measurement performed by an experimentalist with access to subsystem $A$ but not $B$. We define the relevant notions of local and non-local changes in state.
\begin{definition}
Suppose operator $\mathcal{O}$ is normalized in a state $\ket{\Psi}$ so that $\braket{\Psi|\mathcal{O}^\dagger \mathcal{O}|\Psi} = 1$. Suppose also that there exists an operator $\mathcal{O}$(B) local in $\mathcal{H}_B$ such that
\begin{equation}
\braket{\Psi|\mathcal{O}^\dagger \mathcal{O}(B) \mathcal{O}|\Psi} 
\neq
\braket{\Psi|\mathcal{O}(B)|\Psi}.
\end{equation}
If for all $\mathcal{O}(A)$ local in $\mathcal{H}_A$,
\begin{equation}
\braket{\Psi|\mathcal{O}^\dagger \mathcal{O}(A) \mathcal{O}|\Psi} 
=
\braket{\Psi|\mathcal{O}(A)|\Psi},
\end{equation}
then $\mathcal{O}$ changes the state $\ket{\Psi}$ locally in $\mathcal{H}_B$. Otherwise, $\mathcal{O}$ changes the state non-locally.
\label{def:def2}
\end{definition}
It should be understood that when we assess locality by expectation values of operators, we are considering only operators that correspond to observables. We can also assess locality with the reduced density matrix $\rho_A$. If acting with $\mathcal{O}$ does not change $\rho_A$, in other words
\begin{equation}
\rho_A(\mathcal{O}\ket{\Psi}) = \rho_A(\ket{\Psi}),
\end{equation}
then $\mathcal{O}$ changes the state locally in $\mathcal{H}_B$. Local is a special case of localized. A localized operator is local in more than one Hilbert space. 

The definitions we have given for local operators and changes in state apply to field theory. So-called local operators in field theory are local in the quantum-mechanical sense we have defined. Operator $\mathcal{O}(x)$ is local in $\mathcal{H}(x)$, the Hilbert space of the theory restricted to point $x$. When referring to local operators $\mathcal{O}(x)$, we will refer to the point $x$ as the ``insertion point'' of $\mathcal{O}$. The insertion points of a Wilson loop are the points along its integration path. An operator inserted at multiple points is non-local. For example $\mathcal{O}(x)\mathcal{O}(y)$ is non-local but localized to $x,y$. 

We define localized excitations in field theories. This definition is the same as definition \ref{def:def1} but we state it using field theory terminology for clarity. First, we define what we mean by an excitation.
\begin{definition}
In field theory, we call $\mathcal{O} \ket{\Psi}$ an excitation of the state $\ket{\Psi}$ with operator $\mathcal{O}$. 
\end{definition}
The following definition of localized excitations is also known as ``strict localization'' \cite{KnightLocalization}.
\begin{definition}
Consider $\mathcal{O}(A)$ inserted in subregion $A$ of a Cauchy surface. The complement of $A$ on the Cauchy surface is subregion $B$. An operator $\mathcal{O}$ creates an excitation that is localized to $B$ if 
\begin{equation}
\braket{\Psi|\mathcal{O}^\dagger \mathcal{O}(A) \mathcal{O} |\Psi} = \braket{\Psi|\mathcal{O}(A)|\Psi} ~~ \forall ~~ \mathcal{O}(A).
\end{equation}
\end{definition}
The definitions we provided extend in an obvious way to describe operators and excitations localized to a region of spacetime, rather than just a region of a Cauchy surface. A local excitation is an excitation that is localized to a single point. We will sometimes refer to local and non-local excitations of the state in quantum mechanics if we make statements that apply to both quantum mechanics and field theory. Various statements we will make also apply in a natural way to non-localized operators, which are operators inserted in every point in a Cauchy surface. 

In gauge theories, we must fix a gauge before checking the above condition, or we may simply work in terms of gauge-invariant operators. We must also fix a gauge in order to use the reduced density matrix $\rho_A$ to diagnose locality as the density matrix is not gauge-invariant. In this work, we will rarely mention these subtleties involved with gauge theories, as our statements can often be extended in an obvious way to these theories.

The ability to change a state through a non-local excitation should not be confused with the inability for localized experimentalists to transmit information between spacelike-separated entangled systems by performing local measurements. We will explain how these two features of locality are different and consistent in section 5.3.

We state the Reeh-Schlieder theorem, a theorem in quantum field theory that is important for understanding locality considerations. Consider the set of all operators $\mathcal{O}(B)$ that are localized to an open subregion $B$ of a Cauchy surface. These operators generate an algebra $\mathcal{A}(B)$ of the subregion $B$. The complement of $B$ on the Cauchy surface is $B^c$. Suppose the Hilbert space of the theory on the full Cauchy surface is $\mathcal{H}$. Consider the vacuum state of some quantum field theory $\ket{\Omega}$. The Reeh-Schlieder theorem is that states $\mathcal{A}(B) \ket{\Omega}$ are dense in $\mathcal{H}$ \cite{ReehSchlieder}. In other words, one can act with operators that are localized to $B$ to change the state in $B^c$. Moreover, acting with operators localized in $B$ can prepare a state in $B^c$ that is arbitrarily close to any state in $\mathcal{H}$ even if that state is an excitation localized entirely in $B^c$. The Reeh-Schlieder theorem is paradoxical if one assumes that any state $\mathcal{O}(B)\ket{\Omega}$ in principle represents the action of an experimentalist localized to region $B$ on the state $\ket{\Omega}$. The Reeh-Schlieder theorem holds for states other than the vacuum as well. Standard references to the Reeh-Schlieder theorem, as well as other aspects of locality in algebraic and axiomatic quantum field theory include refs. \cite{Haag, StreaterWightman}.

\section{Background: Real-time Perturbation Theory}

We review real-time perturbation theory, otherwise known as the in-in formalism  \cite{InInSchwinger, InInKeldysh}. The formalism is called in-in in contrast to perturbation theory for S-matrix elements, which can be called in-out perturbation theory as it calculates transition amplitudes between initial and final states. Real-time perturbation theory calculates correlation functions 
\begin{equation}
\braket{\Psi| \mathcal{O}_1 \ldots \mathcal{O}_n |\Psi}
\end{equation}
perturbatively in an interaction Hamiltonian with arbitrary time dependence and makes aspects of locality and causality manifest. Beginning with some initial time-independent Hamiltonian $H_0$ and initial state $\ket{\Psi(t_0)}$, time evolution commences, and an interaction Hamiltonian may be turned on. The calculations proceed purely in Lorentzian signature, but the initial state $\ket{\Psi(t_0)}$ may be prepared in various standard ways including by Euclidean path integral. The real-time formalism is also known as the closed-time or Keldysh formalism because the same calculation can be performed using a path integral with a closed-time (Keldysh) contour. The real-time formalism is an inherent part of cosmology and AdS/CFT \cite{Mathur93,Weinberg05, Skenderis08}. 

Just as in-out perturbation theory may be obtained from a Euclidean path integral through Wick rotation, real-time perturbation theory may be obtained from the same Euclidean path integral by deforming the purely imaginary-time contour into a closed-time contour. Calculations proceed similarly for in-out and real-time perturbation theory, both in the use of Feynman diagrams and the treatment of divergences.

To illustrate how real-time perturbation theory works, we calculate the expectation value of Heisenberg-picture operator $\mathcal{O}(t,\mathbf{x})$ in a theory with a time-dependent interaction Hamiltonian $H_{\text{int}}(t)$. We work in $(d+1)$-dimensional spacetime throughout this manuscript. In defining the Heisenberg picture, we use reference time $t_0$. The associated Schrodinger-picture operator defined at time $t_0$ is $\mathcal{O}(t_0,\mathbf{x})$. The full Hamiltonian is 
\begin{equation}
H(t) = H_0 + H_{\text{int}}(t),
\end{equation}
where $H_0$ is time-independent. We use a perturbation that is zero at time $t_0$:
\begin{equation}
H_{\text{int}}(t_0) = 0.
\end{equation} 
We now work in the interaction picture, denoting interaction-picture operators with a subscript $I$. The interaction picture is defined in terms of Schrodinger-picture states and operators as
\begin{align}
\ket{\Psi_I(t)} &= e^{iH_0(t-t_0)}\ket{\Psi(t)}.
\\
\mathcal{O}_I(t,\mathbf{x}) &= e^{iH_0(t-t_0)} \mathcal{O} (t_0,\mathbf{x}) e^{-iH_0(t-t_0)}.
\end{align}
The interaction Hamiltonian $H_{\text{int}}(t)$ in the interaction picture is $H_I(t)$. As we are working in real time, time evolution is unitary and preserves the norm of the state, which we choose to be $\braket{\Psi(t_0)|\Psi(t_0)} = 1$. The time evolution operator $\mathcal{U}(t,t_0)$ is
\begin{equation}
\mathcal{U}(t,t_0) = \mathcal{T}\left( e^{-i\int_{t_0}^t dt' H(t') }\right).
\end{equation}
The interaction-picture evolution operator is
\begin{equation}
\mathcal{U}_I(t,t_0) = \mathcal{T}\left( e^{-i\int_{t_0}^t dt' H_I(t')  }\right)  = e^{iH_0(t-t_0)} \mathcal{U}(t,t_0).
\end{equation}
We may now calculate $\braket{\Psi(t_0)|\mathcal{O}(t,\mathbf{x})|\Psi(t_0)}$ perturbatively in $H_I$. Explicitly,
\begin{align*}
\braket{\Psi(t_0)|\mathcal{O}(t,\mathbf{x})|\Psi(t_0)}
=&
\braket{\Psi(t_0)|\mathcal{U}^\dagger (t,t_0) \mathcal{O} (t_0,\mathbf{x}) \mathcal{U}(t,t_0)|\Psi(t_0)}
\\
=& \bra{\Psi(t_0)} \left(\mathcal{U}^\dagger (t,t_0) e^{-iH_0(t-t_0)} \right)
\left(
e^{iH_0(t-t_0)} \mathcal{O} (t_0,\mathbf{x}) e^{-iH_0(t-t_0)}
\right)
\\
&~~~~~~~~~~~~~~~~~\times
\left(e^{iH_0(t-t_0)}
\mathcal{U}(t,t_0)\right)
\ket{\Psi(t_0)}
.
\numberthis
\end{align*}
Passing into the interaction picture,
\begin{equation}
\braket{\Psi(t_0)|\mathcal{O}(t,\mathbf{x})|\Psi(t_0)}
=\braket{\Psi(t_0)| \mathcal{U}^\dagger_I (t,t_0) \mathcal{O}_I(t,\mathbf{x}) \mathcal{U}_I(t,t_0)|\Psi(t_0)}.
\end{equation}
Expanding in $H_I$,
\begin{align*}
\braket{\Psi(t_0)|\mathcal{O}(t,\mathbf{x})|\Psi(t_0)}
&=
\braket{\Psi(t_0)|\mathcal{O}(t_0,\mathbf{x})|\Psi(t_0)}
+
i \int_{-\infty}^t dt_1 \braket{\Psi(t_0)|[H_I(t_1),\mathcal{O}_I(t,\mathbf{x})]|\Psi(t_0)}
\\
&
-
\int_{-\infty}^t dt_1 \int_{-\infty}^{t_1}  dt_2 
\braket{\Psi(t_0)|[H_I(t_2),[H_I(t_1),\mathcal{O}_I(t,\mathbf{x})]]|\Psi(t_0)}
+\ldots .
\numberthis
\end{align*}
The all-order expression for real-time perturbation theory is given by Weinberg \cite{Weinberg05}.
\begin{align*}
\braket{\Psi(t_0)|\mathcal{O}(t,\mathbf{x})|\Psi(t_0)} =& \sum_{N=0}^{\infty} i^N 
\int_{-\infty}^t dt_N 
\int_{-\infty}^{t_N} dt_{N-1} 
\ldots
\int_{-\infty}^{t_2} dt_1
\\
&~~~~~~~~~~~~
\times \braket{\Psi(t_0)|
\left[
H_I(t_1),
\left[
H_I(t_2),
\left[\ldots
\left[H_I(t_N),\mathcal{O}_I(t,\mathbf{x})\right]
\ldots
\right]
\right]
\right]
|\Psi(t_0)}
.
\numberthis
\label{eqnWeinberg}
\end{align*}
The interaction-picture operators are the Heisenberg-picture operators of the unperturbed theory at time $t_0$.

Real-time perturbation theory makes manifest how turning on a localized interaction creates a localized excitation. Suppose the interaction Hamiltonian is given by some local operator $\mathcal{O}'$ smeared over a spatial region:
\begin{equation}
H_I(t') = \int d^d y f(t',\mathbf{y}) \mathcal{O}_I'(t',\mathbf{y}).
\label{LocalizedHamiltonian}
\end{equation}
The interaction $H_I(t')$ can only change $\braket{\mathcal{O}(t,\mathbf{x})}$ if $f(t',\mathbf{y})$ has support on points that are null or time-like separated from the point $(t,\mathbf{x})$. Only the perturbations for which $t'\leq t$ contribute. If $f(t',\mathbf{y})$ has support only at points spacelike-separated from the point $(t,\mathbf{x})$, then
\begin{equation}
[H_I(t'),\mathcal{O}_I(t,\mathbf{x})] = 0.
\end{equation}
Each term in \eqref{eqnWeinberg} will also vanish for the same reason.

Suppose $H_0$ is a free Hamiltonian and $\ket{\Psi(t_0)} = \ket{0}$, the vacuum of $H_0$. At each order in perturbative expansion, the nested commutators will produce various contractions multiplied by an overall retarded Green's function $G_R(x-y)$, which has precisely the correct causality properties. Diagrammatic rules that make retarded Green's functions manifest are discussed in ref. \cite{Musso06}. In practice, one can obtain the different terms in \eqref{eqnWeinberg} from different analytic continuations of the appropriate Euclidean correlators. The structure of real-time perturbation theory and the presence of retarded Green's functions parallels a problem in classical field theory, calculating corrections to the value of a free field perturbatively in a source.

\section{A Local Picture for Time-Dependent Quantum Systems}

We present a coherent picture for time-dependent operations on pure states in quantum systems. We will refer to this picture as the ``local picture''. This picture is generated by the assumptions that all physical interactions occur through terms in the Hamiltonian, and localized experimentalists deform the Hamiltonian in a localized region. We define the local picture because it unites several different manifestations of locality and causality into one concrete framework. The local picture will provide simple explanations for results in later sections.

We first briefly review the two types of systems we may consider in quantum mechanics. Closed quantum systems are pure states that undergo unitary time evolution. For example, the Hilbert space of a closed quantum system $\mathcal{H}$ may be a tensor-product Hilbert space of a system, experimentalist, and environment:
\begin{equation}
\mathcal{H} = \mathcal{H}_{\text{env}} \otimes \mathcal{H}_{\text{exp}} \otimes \mathcal{H}_{\text{sys}}.
\end{equation}
The experimentalist can be described as an observer and an interaction apparatus:
\begin{equation}
\mathcal{H}_{\text{exp}} = \mathcal{H}_{\text{obs}} \otimes \mathcal{H}_{\text{app}}.
\end{equation}
Open quantum systems are systems that can be acted upon by some external experimentalist. Pure states of an open quantum system are elements of the Hilbert space $\mathcal{H}_{\text{sys}}$. In closed quantum systems, measurement is described by an interaction term in the Hamiltonian that entangles states between $\mathcal{H}_{\text{exp}}, \mathcal{H}_{\text{sys}}$. This is a unitary process and there is no state collapse. Projecting onto one of the states in $\mathcal{H}_{\text{exp}} \otimes \mathcal{H}_{\text{sys}}$ shows the state that one particular experimentalist has access to. In an open quantum system, this measurement process is modelled by projection operators that implement the collapse of the state, which is the Copenhagen interpretation of quantum mechanics, together with a re-normalization of the state. In principle, an open quantum system can be obtained from a closed quantum system, and the details of this process are the subject of current research. We will assume this well-known description is valid formally. In short, to describe the measurement process without collapse and state re-normalization, we must use the closed quantum system. To describe operations performed on the state by an external experimentalist and calculate expectation values, the open quantum system is the natural choice.

We now state the local picture, which governs the evolution of the pure state $\ket{\Psi(t_0)}$ of some system prepared at time $t_0$. Physical operations on the state are described by deformations of the Hamiltonian. Any norm-preserving operation can be treated as a Hamiltonian deformation, but deforming the Hamiltonian by functions of local operators are natural ways to implement physical operations. Localized experimentalists can only deform the Hamiltonian by localized operators, and therefore only act with localized unitary operators on the state. Operators that create non-localized excitations can only be implemented by non-localized experimentalists. Non-localized experimentalists are experimentalists that have access to the entire Cauchy surface, and should not be confused with experimentalists who may depart from the principles of local quantum field theory. We have given the local picture for open quantum systems, but these principles describe closed quantum systems as well.

We give an example that makes the elements of the local picture concrete and shows how they arise. We work in quantum field theory for convenience. The expectation value of operator $\mathcal{O}(\mathbf{x},t)$ evolves as
\begin{equation}
\braket{\mathcal{O}(\mathbf{x},t)} = \braket{\Psi(t_0)|\mathcal{U}^\dagger(t,t_0)\mathcal{O}(\mathbf{x},t_0)\mathcal{U}(t,t_0)|\Psi(t_0)}.
\end{equation}
In order to describe the effect of some interaction Hamiltonian $H_{int}$, we may pass into the interaction picture. As we reviewed in section 3,
\begin{equation}
\braket{\Psi(t_0)|\mathcal{O}(t,\mathbf{x})|\Psi(t_0)}
=\braket{\Psi(t_0)| \mathcal{U}^\dagger_I (t,t_0) \mathcal{O}_I(t,\mathbf{x}) \mathcal{U}_I(t,t_0)|\Psi(t_0)}
\end{equation}
The above expresion is equivalent to the following calculation, in the Heisenberg picture defined by evolution from $t_0$ with Hamiltonian $H_0$:
\begin{align*}
\braket{\mathcal{O}(t,\mathbf{x})} &= \braket{\Psi_e|\mathcal{O}(t,\mathbf{x})|\Psi_e},
\\
\ket{\Psi_e} &\equiv \mathcal{T}\left( e^{-i\int_{t_0}^t dt' \mathcal{H}_{int}(t')} \right) \ket{\Psi(t_0)}.
\numberthis
\end{align*}
We use $\mathcal{H}_{int}$ to denote $H_{int}$ in the Heisenberg picture. It is therefore natural that localized unitary operators $\mathcal{U}$ of the form 
\begin{equation}
\mathcal{U} = \mathcal{T}\left( e^{-i\int_{t_0}^t dt' \mathcal{H}_{int}(t')} \right)
\end{equation}
create localized excitations, and this follows from real-time perturbation theory. This conclusion is independent of perturbation theory, as we will show in Section 6.

As we have shown, there is a correspondence between excitations of the state and Hamiltonian deformations. We focus on localized unitary operators, but this correspondence holds for non-localized unitary operators and their associated non-localized Hamiltonian deformations in the same way. Localized Hamiltonian deformations take the form of local operators smeared over some compact spacetime region, as in equation \eqref{LocalizedHamiltonian}, while for non-localized Hamiltonian deformations the smearing function has support on all of spacetime. It is of course not obvious how to find an explicit $\mathcal{H}_{int}$ or unitary operator $\mathcal{U}$ to represent the action of an arbitrary operator $\mathcal{N} \mathcal{O}$ on a state. Here $\mathcal{N}$ is the state-dependent normalization constant $|\mathcal{N}|^2 = 1/\braket{\Psi(t_0)|\mathcal{O}^\dagger \mathcal{O}|\Psi(t_0)}$.

We comment on what an experimentalist cannot easily do according to the principle that she may only interact with the system through $\mathcal{H}_{int}$. At any given time, she may only interact with the state through operators evaluated at that time, and interactions that last for some finite time must be time-ordered. Other operations, while mathematically valid, are not as natural. For example, it is natural to act with the operator $\mathcal{U}_1$ but not $\mathcal{U}_2$:
\begin{align*}
\mathcal{U}_1 &= \mathcal{T}\left( e^{-i \int dt \mathcal{O}(t) } \right) 
\\
\mathcal{U}_2 &= e^{-i \int dt \mathcal{O}(t) } 
\end{align*}
Calculating correlation functions of operators $\mathcal{O}(t)$ at time $t$ represents experiments conducted at $t$, and excitation of the state that occur after this time do not contribute. These naturalness conditions for operator excitations follow automatically from the local picture.

The local picture reveals that localized experimentalists cannot act with operators that create non-localized excitations. If the experimentalist is localized to some spacelike region, she can only use $\mathcal{H}_{int}$ also localized in this region, which means acting with unitary operators localized to that same region. These operators create localized excitations. In sections 5 and 6, we will find that local non-unitary operators $\mathcal{N} \mathcal{O}$ can create non-localized excitations, and so a localized experimentalist cannot act with these operators. In fact, if $\mathcal{O}$ creates a non-local excitation, the experimentalist must know the state on the entire Cauchy surface in order to calculate $\mathcal{N}$. Operator $\mathcal{N} \mathcal{O}$ can only be acted on the state by a non-localized experimentalist. An example of such an operator is a normalized projection operator that implements a measurement, but to discuss locality in the context of measurements, one is using a closed quantum system either implicitly or explicitly. Projection operators in an open quantum system are simply models of the process. We will discuss measurements explicitly in a later section.

\section{Localized Unitary and Non-unitary Operators in Quantum Mechanics}

We discuss unitary and non-unitary operators in quantum mechanics. Locality properties of operators depend on whether or not they are unitary. Our conclusions in quantum mechanics apply to quantum field theory as well. For a Hilbert space $\mathcal{H} = \mathcal{H}_1 \otimes \mathcal{H}_2$, we address whether acting with an operator local in $\mathcal{H}_1$ may affect expectation values taken in $\mathcal{H}_2$. In this section, we will use a two-particle system of spin $1/2$ particles, where the particles are prepared in product and entangled states. We label the two spin states as $\ket{\pm}$. 

Non-unitary operators generically do not preserve the normalization of states, so to represent their action on the state, we must include a normalization factor along with each operator. This normalization factor must be state-dependent, and so in general non-unitary operators are state-dependent. We will refer to these norm-preserving non-unitary operators as non-unitary operators for short.

Consider an operator local in $\mathcal{H}_2$: 
\begin{equation}
\mathcal{O} = \mathbb{I}_1 \otimes \mathcal{O}_2.
\end{equation}
We will act with $\mathcal{O}$ on different states and calculate the reduced density matrix of particle 1, $\rho_1$. If $\rho_1$ changes, $\mathcal{O}$ has changed the state non-locally. Following the examples, we will prove various general results. The proofs are elementary, and we use elementary methods in order to make certain properties explicit.

We will refer to product states and entangled states of, for example, $\mathcal{H}$. A state $\ket{\Psi} \in \mathcal{H}$ is a product state if there exist states $\ket{\Psi_1} \in \mathcal{H}_1, \ket{\Psi_2} \in \mathcal{H}_2$ such that $\ket{\Psi} = \ket{\Psi_1} \otimes \ket{\Psi_2}$. Product states are also known as separable states. A pure state that is not a product state is entangled.

\subsection{Local operators create local excitations in product states}

In this section, we show how both local unitary and non-unitary operators change product states locally. We show an example and then prove this statement. Choose $\mathcal{O}_2$ to be diagonal for convenience:
\begin{equation}
\mathcal{O}_2 =
\mathcal{N}
\left(
\begin{matrix}
a & 0 \\
0 & b 
\end{matrix} 
\right)
.
\end{equation}
Here, $a,b \in \mathbb{C}$. If $a,b$ are pure phases then $\mathcal{O}_2$ is unitary. Here $\mathcal{N}$ is a normalization factor. 

First, consider the product state 
\begin{equation}
\ket{\Psi^p} = \ket{+}_1 \ket{-}_2
.
\end{equation}
The reduced density matrix $\rho^p_1$ is
\begin{equation}
\rho_1^p = \ket{+}\bra{+}.
\end{equation}
Acting with $\mathcal{O}$,
\begin{equation}
 \mathcal{O} \ket{\Psi^p} = \mathcal{N} \ket{+}_1( b \ket{-}_2) \equiv \ket{\Psi^{p'}}.
\end{equation}
To normalize the state, $|\mathcal{N}|^2 = 1/|b|^2$. The reduced density matrix is unchanged:
\begin{equation}
\rho_1^{p'} = \ket{+} \bra{+}.
\end{equation}
Acting with $\mathcal{O}$ does not change measurements performed on particle 1 regardless of what values $a,b$ take.

We now prove that all local operators, including non-unitary operators, change product states locally. Consider two Hilbert spaces $\mathcal{H}_{1,2}$ and $n$ orthonormal basis elements $\ket{\psi_{1,2}^n}$. Consider the arbitrary product state $\ket{\Psi_p} \in \mathcal{H}$ with $\mathcal{H} = \mathcal{H}_1 \otimes \mathcal{H}_2$, and an arbitrary norm-preserving operator $\mathcal{O}$ local in $\mathcal{H}_2$.
\begin{equation}
\ket{\Psi_p} = \sum_i c_1^i \ket{\psi_1^i} \sum_j c_2^j \ket{\psi_2^j}.
\end{equation}
The state is normalized:
\begin{equation}
\left( \sum_i |c_1^i|^2 \right)  \times \left( \sum_j |c_2^j|^2 \right) = 1.
\end{equation}
The reduced density matrix associated with $\mathcal{H}_1$ is
\begin{equation}
\rho_1 = \left(\sum_i c_1^i \ket{\psi_1^i} \right) 
\left(
\sum_k
c_1^{k*} \bra{\psi_1^k} \right) \sum_j |c_2^j|^2.
\end{equation}
Acting with $\mathcal{O}$ on the state,
\begin{equation}
\mathcal{O}\ket{\Psi} = \sum_i c_1^i \ket{\psi_1^i} \sum_j d_2^{j} \ket{\psi_2^j},
\end{equation}
where the $d_2^{j}$ are defined by the action of $\mathcal{O}$ on states in $\mathcal{H}_2$ in the chosen basis: $d_2^j \equiv \sum_i \mathcal{O}_{ji} c_2^i $. The normalization condition is
\begin{equation}
\left( \sum_i |c_1^i|^2 \right)  \times \left( \sum_j |d_2^j|^2 \right) = 1.
\end{equation}
Acting with $\mathcal{O}$ is a unitary operation in $\mathcal{H}_2$:
\begin{equation}
\sum_j |d_2^j|^2  = \sum_j |c_2^j|^2 .
\end{equation}
We may see that the new reduced density matrix $\rho_1'$ is equal to $\rho_1$:
\begin{align*}
\rho_1' &= \left(\sum_i c_1^i \ket{\psi_1^i} \right) 
\left(
\sum_k
c_1^{k*} \bra{\psi_1^k} \right) \sum_j |d_2^j|^2
\\
&=\left(\sum_i c_1^i \ket{\psi_1^i} \right) 
\left(
\sum_k
c_1^{k*} \bra{\psi_1^k} \right) \sum_j |c_2^j|^2 =\rho_1 \numberthis .
\end{align*}
This concludes the proof.

\subsection{Locality in entangled states}

In entangled states, local unitary operators affect the state locally but local non-unitary operators may affect the state non-locally. As an example, we will act with $\mathcal{O}$ on entangled state $\ket{\Psi_e}$ and find that generically $\mathcal{O}$ will change the state non-locally unless $\mathcal{O}_2$ is unitary.
\begin{equation}
\ket{\Psi^e} = \frac{1}{\sqrt{2}}(\ket{+}_1 \ket{-}_2 - \ket{-}_1\ket{+}_2).
\end{equation}
The reduced density matrix $\rho_1^e$ is
\begin{equation}
\rho_1^e = \frac{1}{2}(\ket{+} \bra{+} + \ket{-} \bra{-}).
\end{equation}
Acting with $\mathcal{O}$ on the state $\ket{\Psi^e}$ gives
\begin{equation}
\mathcal{O} \ket{\Psi^e} = \frac{\mathcal{N}}{\sqrt{2}}(b \ket{+}_1 \ket{-}_2 - a \ket{-}_1\ket{+}_2) \equiv \ket{\Psi^{e'}}.
\end{equation}
The normalization factor satisfies $|\mathcal{N}|^2 = \frac{2}{|a|^2 + |b|^2}$. The reduced density matrix is now
\begin{equation}
\rho_1^{e'} = \frac{1}{|a|^2+|b|^2}(|b|^2 \ket{+} \bra{+} + |a|^2 \ket{-} \bra{-}).
\end{equation}
Acting with $\mathcal{O}$ changes $\rho_1^e$ unless $|a|^2 = |b|^2$, which would make $\mathcal{O}_2$ unitary.

Local operators mix states within $\mathcal{H}_2$, but if these states are coupled with different weights to states in a different Hilbert space $\mathcal{H}_1$, mixing states within $\mathcal{H}_2$ will generically change the relative weights of the states in $\mathcal{H}_1$. Only local unitary operators mix states in $\mathcal{H}_2$ in the way that leaves the partial trace unchanged.

The action of any non-unitary operator $\mathcal{N} \mathcal{O}$ on a state may by definition be written as a unitary operator $\mathcal{U}$ acting on the state. In the state $\ket{\Psi}$, $|\mathcal{N}|^2 = \braket{\Psi|\mathcal{O}^\dagger \mathcal{O}|\Psi}^{-1}$. For every $\mathcal{O}$ and $\ket{\Psi}$ there exists a unitary operator $\mathcal{U}$ such that
\begin{equation}
\mathcal{N} \mathcal{O} \ket{\Psi} = \mathcal{U}\ket{\Psi}.
\end{equation}
This equality follows from the fact that acting with $\mathcal{N} \mathcal{O}$ does not change the norm of the state. It follows that this norm-preserving operation on a state can be implemented by acting with some unitary operator $\mathcal{U}$, as the set of all unitary operators is the space of all possible norm-preserving operations on the state. The non-unitary operator $\mathcal{N} \mathcal{O}$ is of course not equal to the corresponding unitary operator $\mathcal{U}$, but their actions on the state $\ket{\Psi}$ are the same. The same $\mathcal{O}$ is represented by different $\mathcal{U}$ on different states $\ket{\Psi}$. If the operator $\mathcal{O}$ changes the state non-locally, $\mathcal{U}$ is non-local. The equivalence between non-unitary and unitary operators acting on the state shows how non-unitary operators may be applied to a state through time evolution. According to the local picture, the unitary operator $\mathcal{U}$ is applied by a non-local experimentalist with access to both systems, as $\mathcal{U}$ is not local in $\mathcal{H}_1 $ or $\mathcal{H}_2$ alone.

Let us see an explicit example of the equivalence between non-unitary and unitary operators using the operator $\mathcal{O}$ and state $\ket{\Psi^e}$. We now work in the basis of the full Hilbert space: $(\ket{+}\ket{+},\ket{+} \ket{-}, \ket{-} \ket{+},\ket{-} \ket{-})$ where we have dropped the subscripts $1, 2$. We wish to find a unitary operator $\mathcal{U}$ that satisfies the following:
\begin{equation}
\mathcal{U} \ket{\Psi^e} = \mathcal{O} \ket{\Psi^e}.
\end{equation}
Writing this condition in the chosen basis,
\begin{equation}
\mathcal{O} \ket{\Psi} =
\mathcal{N} \frac{1}{\sqrt{2}}
\left(
\begin{matrix}
0 \\
b \\
-a \\
0
\end{matrix}
\right)
=
\mathcal{U} \frac{1}{\sqrt{2}}
\left(
\begin{matrix}
0 \\
1 \\
-1 \\
0
\end{matrix}
\right) = \mathcal{U} \ket{\Psi}.
\end{equation}
We can now write down a solution. $\mathcal{U}$ rotates components into one another, and may produce an arbitrary phase.
\begin{equation}
\mathcal{U}(\theta,\phi_1,\phi_2) = 
\left(
\begin{matrix}
1 & 0 & 0 & 0 \\
0 & e^{i\phi_1}\cos\theta & -e^{i\phi_2}\sin\theta & 0 \\
0 & e^{i\phi_1}\sin\theta & e^{i\phi_2}\cos\theta & 0 \\
0 & 0 & 0 & 1
\end{matrix}
\right).
\end{equation}
The above matrix is the product of the rotation matrix with $\text{diag}(1,e^{i\phi_1},e^{i\phi_2},1)$. The relations between the angles and $a,b$ are:
\begin{align}
e^{i \phi_1} \cos \theta  + e^{i \phi_2} \sin \theta= b \mathcal{N}
\\
e^{i\phi_1} \sin \theta - e^{i \phi_2} \cos \theta = -a \mathcal{N}.
\end{align}
Elements $a\mathcal{N},b\mathcal{N}$ have three degrees of freedom: an overall phase, a relative phase, and a relative magnitude. Operator $\mathcal{U}$ has the same three degrees of freedom as well: two phases $\phi_1, \phi_2$, the angle $\theta$ that controls the components' magnitudes. The normalization condition $(|a\mathcal{N}|^2+|b\mathcal{N}|^2)/2 = 1$ is satisfied. We see that the unitary operator $\mathcal{U}$ depends on $\mathcal{O}$ and the state.

We will now show that local unitary operators change entangled states locally and generic local non-unitary operators change entangled states non-locally. This is an elementary property of the partial trace, but we will find a proof in component notation useful. We begin with an arbitrary entangled state $\ket{\Psi_e}$. We use repeated index summation notation in this proof. Label states in $\mathcal{H}_{1,2}$ by $\ket{\psi_{1,2}}$.
\begin{equation}
\ket{\Psi_e} = C_{ia} \ket{\psi_1^i} \ket{\psi_2^a}.
\end{equation}
The normalization condition is $C_{ai}^\dagger C_{ai} = 1$. The reduced density matrix of $\mathcal{H}_1$ is
\begin{equation}
\rho_1 =  C_{ia} C_{aj}^\dagger\ket{\psi_1^i} \bra{\psi_1^j}.
\end{equation}
Now act with an operator $\mathcal{O}$ that is local in $\mathcal{H}_2$ on the state.
\begin{equation}
\ket{\Psi_e} = \mathcal{O}_{ba} C_{ia} \ket{\psi_1^i} \ket{\psi_2^b}.
\end{equation}
Assume that $\mathcal{O}$ is normalized to satisfy the normalization condition $ C_{di}^\dagger \mathcal{O}_{db}^\dagger \mathcal{O}_{ba} C_{ia}= 1$. If $\mathcal{O}^\dagger = \mathcal{O}^{-1}$ then $\mathcal{O}_{db}^\dagger \mathcal{O}_{ba} = \mathbb{I}_{da}$ and the state's norm is automatically preserved. The new reduced density matrix $\rho_1'$ is
\begin{equation}
\rho_1' = C_{dj}^\dagger \mathcal{O}_{db}^\dagger \mathcal{O}_{ba} C_{ai} \ket{\psi_1^i}\bra{\psi_1^j}.
\end{equation}
If operator $\mathcal{O}$ is unitary, then $\rho_1' = \rho_1$. If not, the state may change non-locally. In the language of quantum information, non-unitary operators $\mathcal{O}$ implement non-local quantum gates, as $\mathcal{O}$'s action can be represented by a non-local unitary operator.

There are states for which non-unitary operators acting on a subspace do leave $\rho_1$ unchanged. Suppose that for $a>k$, $C_{ai} = 0 $ for every $i$. To leave $\rho_1$ unchanged,  $(\mathcal{O}^\dagger \mathcal{O})_{db} = \mathbb{I}_{db}$ for $d,b \leq k$ suffices, but this condition not necessary for $d,b > k$. A simple example is if $\mathcal{O}$ is local and unitary in subspace $P \in \mathcal{H}$ but non-unitary in the rest of $\mathcal{H}$: $\mathcal{O}$ will create a local excitation of states in $P$ despite being non-unitary. When the two Hilbert spaces have the same dimensionality, such states require the density matrix to have at least one zero eigenvalue.

We have not addressed the most general condition for operators to leave entanglement entropy unchanged, which is a weaker condition than leaving $\rho_1$ unchanged, and an interesting direction for future work. Entanglement entropy is the von Neumann entropy of a reduced density matrix. The entanglement entropy $S_1$ of the subsystem with Hilbert space $\mathcal{H}_1$ is
\begin{equation}
S_1 = - \text{tr}(\rho_1 \ln \rho_1)
\end{equation}
In a pure state, entanglement entropies for the two subsystems must be equal: $S_1 = S_2$. It follows that acting with a local unitary operator on one subsystem or the other does not change $S_1, S_2$ because the density matrix of the subsystem not acted upon is unchanged.

There can be local non-unitary operators $\mathcal{O}$ that alter a state $\ket{\Psi}$ non-locally, but leave expectation values of operators $\mathcal{O}'$ unchanged. In field theory this condition is sometimes satisfied by the modular Hamiltonian, $\mathcal{O}' = -\ln(\rho_1)$, whose expectation value is entanglement entropy \cite{NozakiNT14}. It would be interesting to investigate this question further in quantum mechanics and field theory to understand the basis of the apparent quasi-particle picture for entanglement entropy.

\subsection{Causality, non-local state preparation, and Reeh-Schlieder}

So far, we have shown how local non-unitary operators act as unitary operators on entangled states, and that these unitary operators must be non-local. According to the local picture, operators that change states non-locally can only be implemented by non-local experimentalists. Experimentalists act with the non-local unitary operators through time evolution. Non-unitary operators are intrinsically non-local in entangled states.

Our results can be viewed another way, motivated by the Reeh-Schlieder theorem in field theory. In a state in $\mathcal{H} = \mathcal{H}_1 \otimes \mathcal{H}_2$ that is entangled between $\mathcal{H}_1,\mathcal{H}_2$, non-unitary operators $\mathcal{O}$ local in $\mathcal{H}_2$ can prepare states in $\mathcal{H}_1$. We may see how this works for $\ket{\Psi^e}$.
\begin{equation}
\ket{\Psi^e} = \frac{1}{\sqrt{2}}(\ket{+}_1 \ket{-}_2 - \ket{-}_1\ket{+}_2).
\end{equation}
We may use suitably normalized operators $L_\pm, L_z$ acting on particle 2 that are the angular momenta operators for spin $1/2$ particles. Ignoring the overall normalizations,
\begin{align*}
L_+ \ket{\Psi^e}&= \ket{+}_1\ket{+}_2.
\\
L_- \ket{\Psi^e}&= \ket{-}_1\ket{-}_2.
\\
(1+L_z) \ket{\Psi^e}&= \ket{-}_1\ket{+}_2.
\\
(1-L_z) \ket{\Psi^e}&= \ket{+}_1\ket{-}_2.
\numberthis
\end{align*}

Our results show that seemingly counter-intuitive features of the Reeh-Schlieder theorem are perfectly straightforward in quantum mechanics. The fact that local non-unitary operators create non-local excitations in entangled states is the origin of the non-local state preparation in the Reeh-Schlieder theorem. The theorem is consistent with causality because local experimentalists act only with local unitary operators, which do not permit non-local state preparation. The example we gave of preparing a state non-locally was used as a quantum-mechanical model for the Reeh-Schlieder theorem in the leading interpretation \cite{Redhead95}. Discussions of the Reeh-Schlieder theorem have previously been centered on non-local state preparation through measurements in open quantum systems. 

Our conclusions about the non-locality of non-unitary operators may appear to contradict a standard statement of causality for entangled states that allows non-local changes in state, but in fact the two are compatible. Measurements in an open quantum system can be described in the associated closed quantum system by projecting onto an experimentalist-system state with a particular measurement outcome. It is obvious that replacing a superposition of states with one of its constituent states is a non-local change of state, but questions of locality are more clearly formulated in the closed quantum system, in which it is manifest how measurement does not change states non-locally. These statements are standard, but for completeness we now make them concrete with an explicit example.

Consider an initial state of a closed quantum system with Hilbert space $\mathcal{H} = \mathcal{H}_{\text{exp}} \otimes \mathcal{H}_{\text{sys}}$,
\begin{equation}
\ket{\Psi} = \ket{\Psi_{\text{exp}}^1} \ket{\Psi_{\text{exp}}^2} (a \ket{+}_1 \ket{-}_2 - b \ket{-}_1 \ket{+}_2).
\end{equation}
We label the experimentalists by the outcome they observe as $\ket{\Psi_{\text{exp}}(\pm)}$. After experimentalist 2 measures the spin of particle 2, the state is
\begin{equation}
\ket{\Psi}' = \ket{\Psi_{\text{exp}}^1}  (a \ket{\Psi_{\text{exp}}^2(-)} \ket{+}_1 \ket{-}_2 - b \ket{\Psi_{\text{exp}}^2(+)} \ket{-}_1 \ket{+}_2).
\end{equation}
Experimentalist 1 may now measure the spin of particle 1. Once again, this is an interaction that couples experimentalist states to system states. The new state is
\begin{equation}
\ket{\Psi}'' = a \ket{\Psi_{\text{exp}}^1(+)}   \ket{\Psi_{\text{exp}}^2(-)} \ket{+}_1 \ket{-}_2 - b \ket{\Psi_{\text{exp}}^1(-)}  \ket{\Psi_{\text{exp}}^2(+)} \ket{-}_1 \ket{+}_2).
\end{equation}
We now could project onto various states to determine what each experimentalist measures. However, tracing out experimentalist 2 and particle 2, the reduced density matrix $\rho_{\text{exp}_1 \otimes \text{sys}_1}$ for experimentalist 1 and particle 1 is
\begin{equation}
\rho_{\text{exp}_1 \otimes \text{sys}_1} = |a|^2 \left( \ket{\Psi_{\text{exp}}^1(+)} \ket{+} \bra{+} \bra {\Psi_{\text{exp}}^1(+)} \right)
+ |b|^2  \left( \ket{\Psi_{\text{exp}}^1(-)} \ket{-} \bra{-}  \bra{\Psi_{\text{exp}}^1(-)} \right).
\end{equation}
The reduced density matrix is unchanged by the measurement that experimentalist 2 performed. The unitary operator that implements experimentalist 2's measurement is localized to $\mathcal{H}_{\text{sys}}^2 \otimes \mathcal{H}_{\text{exp}}^2$, and so it does not change $\rho_{\text{exp}_1 \otimes \text{sys}_1}$.

\subsection{Superpositions of localized excitations}

In entangled states, the superposition of two localized excitations is generically not itself a localized excitation. One can represent each localized excitation as a localized unitary operator acting on some reference state. The sum of two unitary operators need not be unitary, and so their superposition need not create a localized excitation. For example, by adding two unitary matrices of the form $\text{diag}(e^{i\phi_1}, e^{i\phi_2})$, we may change the magnitude of the sum's diagonal entries. The non-locality of the superposition measures a kind of interference between the two unitary operators. We can state the general condition for which the superposition of localized excitations implemented by $\mathcal{U}_1, \mathcal{U}_2$ must itself be a localized excitation in entangled states.
\begin{equation}
\mathcal{U}_1 \mathcal{U}_2^\dagger + \mathcal{U}_2 \mathcal{U}_1^\dagger = 0
\end{equation}
We will not explore this condition. It follows that a local experimentalist cannot superimpose two local excitations of an entangled state. The condition for superpositions of local unitary operators to be local has been addressed on a more formal level \cite{Licht63}.

There is a natural way to combine local excitations without superposition. Acting with two localized unitary operators on the same state creates a localized excitation. For example, the operator $\mathcal{U}_1 \mathcal{U}_2$ creates an excitation localized to the same Hilbert spaces in which $\mathcal{U}_1, \mathcal{U}_2$ are localized. In the local picture, these two operators should be time-ordered: $\mathcal{T}\left( \mathcal{U}_1 \mathcal{U}_2 \right)$. This method of combining local excitations is natural in the local picture, as it corresponds to an interaction that occurs in two subsystems:  $\mathcal{T} \left( e^{i (\mathcal{O}_1 + \mathcal{O}_2)} \right)$ and is implemented by turning on sources for both $\mathcal{O}_1$ and $\mathcal{O}_2$. We will elaborate on this point in a later section when we discuss separable and non-separable localized unitary operators.

Superpositions of local unitary operators create non-local excitations, which are associated more naturally with non-unitary operators. This relationship between unitary and non-unitary operators can be viewed from another direction. Non-unitary operators can be written as superpositions of unitary operators. Acting with a non-unitary operator is equivalent to superimposing states formed by acting with unitary operators. As is standard, any operator $\mathcal{O}$ can be written as a linear combination of an Hermitian operator $H_+$ and anti-Hermitian operator $H_-$. 
\begin{align}
\mathcal{O} &= H_+ + H_-.
\\
H_\pm &= \frac{1}{2} (\mathcal{O} \pm \mathcal{O}^\dagger).
\end{align}
In fact, $H_-$ can be written in terms of a Hermitian operator $H_+'$ simply: $H_+' = i H_-$. We now show that any Hermitian operator can be written in terms of a unitary operator and its adjoint, subject to a certain condition. Suppose the spectrum of some Hermitian operator $H$ is bounded from above and below. This is not always the case for the Hamiltonian, but this is true for many other operators, especially those relevant in systems with a finite number of spins. There exists $\lambda$ which is at least as large as $H$'s largest-magnitude eigenvalue, but finite. Operator $H$ is given by
\begin{equation}
H = \frac{1}{2|\lambda|} (U + U^\dagger).
\end{equation}
To prove this, first suppose $H$ is diagonal. Its entries are its eigenvalues, which are real. We may choose $U = \text{diag}(e^{i\phi_1}, e^{i\phi_2},\ldots)$.
\begin{equation}
U+U^\dagger = \text{diag}(2 \cos(\phi_1), 2 \cos(\phi_2), \ldots).
\end{equation}
One may then choose each $\phi_i$ to match each eigenvalue in $H$. Hermitian matrices are diagonalized by unitary matrices, so we may use unitary $V$ to produce any other Hermitian operator from $H$ that has the same eigenvalues: 
\begin{align}
V H V^\dagger &= \frac{1}{2|\lambda|}(V U V^\dagger + V U^\dagger V^\dagger)
\\
&\equiv \frac{1}{2|\lambda|}(U' + U'^\dagger).
\end{align}
Note that $U'$ is also unitary. This concludes the proof.

\section{Localized Unitary and Non-unitary Operators in Quantum Field Theory}

We now turn to field theory and our main result: localized unitary operators create localized excitations, while more familiar local non-unitary operators generically create non-local excitations. We explore the properties of these operators. As generic states in field theory are entangled over spatial regions, field theory is often the study of operators in entangled states. The properties we found in section 5 will apply in field theory as well. An initial investigation into the locality of certain local unitary operators was conducted in the context of free field theory, and our work extends this investigation \cite{KnightLocalization}.

\subsection{Localized unitary operators create localized excitations}

In a $(d+1)$-dimensional theory, we give a general form for time-ordered localized unitary operators $\mathcal{U}$, or localized unitary operators for short. This form arises naturally in the local picture. Foliate the spacetime by Cauchy surfaces and define a timelike coordinate that parameterizes motion across these surfaces. For any given foliation, all operators of the following form are localized unitary operators:
\begin{equation}
\mathcal{U} = \mathcal{T}\left(  e^{-i\sum_n \left( \prod_{\left\{i_n\right\}} \int_{R_{i_n}}  d^{d+1}  x_{i_n} \right)  J_n(x_1,x_2,\ldots,x_{i_n})  \left( \prod_{\left\{i_n\right\}} \mathcal{O}_{i_n} (x_{i_n}) \right)} \right).
\end{equation}

For each $n$, there is a set denoted by $\left\{i_n\right\}$ which specifies the source function $J_n$ and operators $\mathcal{O}_{i_n}(x_{i_n})$ appearing in the product. Functions $J_n$ can have dimensions and include a small expansion parameter. The term in the exponent multiplying $i$ is Hermitian. Unitary operators can always be placed in exponential form but the expression we present is simply the general time-ordered exponential of products and sums of localized operators with smearing functions. Operators $\mathcal{O}(x_{i_n})$ need not be local in space but must be local in time, and we have labelled operators $\mathcal{O}_{i_n}$ by their insertion points schematically.

$R_{i_n}$ is defined as the spacetime region in which the smearing function $J_n(x_1,\ldots,x_{i_n})$ is non-zero. The localized unitary operator $\mathcal{U}$ creates an excitation localized to spacetime region $R = \cup_{i_n} R_{i_n} $. Correlators of operators inserted at spacelike separation from all points in $R$ do not change. Consider $\braket{\mathcal{O}(y)}$ in an excited state formed by $\mathcal{U}$. If $y$ is spacelike-separated from all points in $R$,
\begin{equation}
\braket{\Psi|\mathcal{U}^\dagger \mathcal{O}(y) \mathcal{U}|\Psi} = \braket{\Psi|\mathcal{U}^\dagger  \mathcal{U} \mathcal{O}(y)|\Psi} = \braket{\Psi|\mathcal{O}(y)|\Psi}.
\end{equation}
If $y$ is not spacelike from all of $R$, the above equality may not hold. While we may find it convenient to use perturbation theory to calculate correlators in this state, this result is true non-perturbatively, and follows from \eqref{eqnCommutator}.

There is an operator-excited state correspondence for subregions. For every Hermitian operator $\mathcal{O}$ inserted in a subregion $A$ of a Cauchy surface, there is an excited state whose excitation is localized to $A$ and is given by acting $e^{i\mathcal{O}}$ on the original state.

Just as $\mathcal{H}_{int}$ is not normal ordered, the operators in the exponent of $\mathcal{U}$ are not normal-ordered. Correlators in this state will generically diverge. Treating $\mathcal{H}_{int}$ as a perturbative correction, calculating correlators in a state created by $\mathcal{U}$ amounts to a calculation in real-time perturbation theory, and the divergences are treated using standard methods. 

A simple example of a local unitary operator is
\begin{equation}
\mathcal{U}(x) = e^{-i\alpha\mathcal{O}(x)}, \quad \mathcal{O}^\dagger(x) = \mathcal{O}(x).
\end{equation}
The parameter $\alpha$ can be chosen to be $\alpha \equiv \alpha' \epsilon$, where $\alpha'$ may have dimensions and the dimensionless parameter $\epsilon$ may be taken small. Exciting a state with this operator is equivalent to introducing the interaction 
\begin{equation}
\int \mathcal{H}_{int} = \alpha\int d^{d+1}x' \delta^{d+1}(x'-x)\mathcal{O}(x').
\end{equation}
The first-order correction to $\braket{\mathcal{O}(y)}$ is a familiar quantity in time-dependent systems, the commutator.
\begin{equation}
\braket{\Psi|\mathcal{U}^\dagger(x) \mathcal{O}(y) \mathcal{U}(x)|\Psi}
=
\braket{\Psi|\mathcal{O}(y)|\Psi} -i\alpha \braket{\Psi|[\mathcal{O}(y),\mathcal{O}(x)]|\Psi} + \ldots.
\end{equation}
Conversely, calculations of the commutator of two operators are also the first-order correction to the one-point function in an excited state. In general, the correspondence between localized unitary operators and Hamiltonian deformations is
\begin{align*}
\mathcal{U} &= \mathcal{T}\left(  e^{-i\sum_n \left( \prod_{\left\{i_n\right\}} \int_{R_{i_n}}  d^{d+1}  x_{i_n} \right)  J_n(x_1,x_2,\ldots,x_{i_n})  \left( \prod_{\left\{i_n\right\}} \mathcal{O}_{i_n} (x_{i_n}) \right)} \right)
\\
&~~~~~~~~~~~~~~~~~~~~~~~~~~~~~~~~~~\updownarrow
\\
\int \mathcal{H}_{int} &= \sum_n \left( \prod_{\left\{i_n\right\}} \int_{R_{i_n}}  d^{d+1}  x_{i_n} \right)  J_n(x_1,x_2,\ldots,x_{i_n})  \left( \prod_{\left\{i_n\right\}} \mathcal{O}_{i_n} (x_{i_n}) \right).
\numberthis
\end{align*}
This correspondence is clear from the interaction picture and section 4.

Deformations of the Hamiltonian cannot always be represented by localized unitary operators acting on states. For example, if $\mathcal{H}_{int}$ has not turned off at the time operators are inserted, the calculation of an unequal-time correlator will involve insertions of operators $e^{-i\int^t \mathcal{H}_{int}}$ between operators inserted at different times. Also, the time $t$ should not be taken later than the latest time at which the operators in the correlation function are inserted. These rules follow from the interaction picture.

In gauge theories, it is natural to restrict the operators in the exponent of a localized unitary operator to be gauge invariant. For example, the operator $\mathcal{U}(x)=e^{iF^2(x)}$ with $F_{\mu \nu}(x)$ being the field strength tensor of a gauge theory creates a local excitation. Not all quantities diagnose locality well in gauge theories. For example, entanglement entropy is not a gauge-invariant measure, but relative entropy and mutual information are free of ambiguities associated with the gauge theories, and so may prove useful \cite{CasiniHR13,Soni15 ,Ma2015}.

Localized excitations can change conserved quantities. For example, a localized excitation created by $\mathcal{U}(x)$ will generically change the total energy $\braket{\int d^d y T^{00}(y) }$. The energy added by $\mathcal{U}(x)$ is injected at $x$ and can spread within the forward lightcone of $x$. The amount by which a localized unitary operator changes a conserved quantity is a property of both the operator and the state. It has been argued that localized excitations of fixed particle number are of limited applicability \cite{KnightLocalization,NewtonWigner,Hegerfeldt74,Halvorson00}.

We may also prepare a localized excitation by sending in an ``ingoing excitation'' rather than by deforming the Hamiltonian. So far, we have described how to create a localized excitation by applying a localized unitary operator $\mathcal{U}$ to some state $\ket{\Psi}$. Applying this operator can change conserved quantities. The same localized excitation can be prepared by beginning with some initial state and evolving time. Conserved quantities will not change in this case. The initial state $\ket{\Psi(t_0)}$ that encodes the ingoing excitation is found by evolving the state $\mathcal{U}(x)\ket{\Psi}$ backwards in time. Here, $\mathcal{U}(t,t_0)$ is the time evolution operator and $\mathcal{U}(x)$ is a local unitary operator inserted at spacetime point $x = (t,\mathbf{x})$.
\begin{equation}
\ket{\Psi(t_0)} = \mathcal{U}(t_0,t)\mathcal{U}(x)\ket{\Psi}.
\end{equation}
Even in an interacting field theory in an arbitrary number of dimensions, the ingoing excitation must be a pulse that leaves no imprint on the state as it passes through a spacetime region, because the regions it passes through are causally connected to points that are spacelike-separated from $x$. While such excitations are familiar in classical theories, in quantum field theories they can require extensive fine-tuning, and may not be possible in practice.

The dynamics of entanglement at different scales within a local excitation can be investigated through the time-dependence of entanglement density \cite{BhattacharyaHRT14}. This investigation may be useful in the AdS/CFT correspondence through the Hubeny-Rangamani-Takayanagi conjecture, and as part of the entanglement tsunami picture \cite{Liu13,Leichenauer15, HubenyRangamaniTakayanagi, RyuTakayanagi}.

\subsection{Separable vs. non-separable localized unitary operators}

There are two qualitatively different types of excitations created by localized unitary operators, separable and non-separable. We explore their properties. We use the labels separable and non-separable for reasons which will become clear. 

We have stated a general form for useful localized unitary operators is
\begin{align*}
\mathcal{U} &= \mathcal{T}\left(  e^{-i\sum_n \left( \prod_{\left\{i_n\right\}} \int_{R_{i_n}}  d^{d+1}  x_{i_n} \right)  J_n(x_1,x_2,\ldots,x_{i_n})  \left( \prod_{\left\{i_n\right\}} \mathcal{O}_{i_n} (x_{i_n}) \right)} \right).
\end{align*}
Separable unitary operators $\mathcal{U}$ create separable excitations, and take the form
\begin{equation}
\mathcal{U} = \mathcal{T}\left(  e^{-i\sum_n  \int_{R_n} d^{d+1} x_n J_n(x_n)\mathcal{O}_n(x_n)} \right).
\end{equation}
Non-separable unitary operators contain products of operators in the exponent that are inserted at different points. For example, consider separable and non-separable local unitary operators $\mathcal{U}_s, \mathcal{U}_{ns}$, with
\begin{align}
\mathcal{U}_s &= e^{-i(\mathcal{O}(x)+\mathcal{O}(y))}.
\\
\mathcal{U}_{ns} &= e^{-i\mathcal{O}(x)\mathcal{O}(y)}.
\end{align}
Points $x,y$ are spacelike-separated and so time-ordering has no effect for these two operators. We require that $\mathcal{O}^\dagger = \mathcal{O}$. Operator $\mathcal{U}_s$ can be separated into the product of two local unitary operators while $\mathcal{U}_{ns}$ cannot. This will become obvious shortly.

We can understand separable and non-separable unitary operators through their quantum-mechanical analogs. Consider an entangled state of two spin 1/2 particles. A separable operator is $\mathcal{U}_s = e^{-i( S_z^{1} + S_z^{2})}$, which amounts to acting with a local unitary operator on each particle. Separable operators are the natural way to concatenate local excitations of a system. A non-separable operator is $\mathcal{U}_{ns} = e^{-i S_z^{1} \otimes S_z^{2}}$. This is equivalent to turning on a spin-spin coupling between the two systems. Separable operators represent interaction of an external system with the state and non-separable operators represent the coupling of two subsystems. In the local picture, separable operators are implemented by a non-local experimentalist.

Separable unitary operators represent uncorrelated localized excitations while non-separable unitary operators represent correlated excitations. We will explore this statement in field theory. If two excitations are correlated, correlators affected by one excitation will also depend on the value of the field at the location of the second excitation. Consider the expectation value of operator $\mathcal{O}(z)$ which is altered by the excitation at $x$ but not $y$. Suppose $z$ and $x$ are timelike-separated and $z$ is in the future of $x$. Point $z$ is spacelike-separated from $y$. For the separable operator,
\begin{equation}
\braket{\Psi|\mathcal{U}_s^\dagger(x,y) \mathcal{O}(z) \mathcal{U}_s(x,y)|\Psi}
=
\braket{\Psi|e^{i\mathcal{O}(x)}\mathcal{O}(z) e^{-i\mathcal{O}(x)}|\Psi}.
\end{equation}
For the non-separable operator,
\begin{align*}
\braket{\Psi|\mathcal{U}_{ns}^\dagger(x,y) \mathcal{O}(z) \mathcal{U}_{ns}(x,y)|\Psi}
&=
\braket{\Psi|\mathcal{O}(z) + i[\mathcal{O}(x) \mathcal{O}(y), \mathcal{O}(z)] +\ldots  |\Psi}
\\
&=\braket{\Psi|\mathcal{O}(z) + i\mathcal{O}(y) [\mathcal{O}(x) , \mathcal{O}(z)] +\ldots  |\Psi}.
\numberthis
\end{align*}
Both separable and non-separable unitary operators create localized excitations, just as $\mathcal{U}_s, \mathcal{U}_{ns}$ change the state only in the forward lightcones of their insertion points $x,y$. With $\mathcal{U}_{ns}$, the correction to $\braket{\mathcal{O}(z)}$ depends on an operator $\mathcal{O}(y)$ inserted at spacelike separation. Just as in quantum mechanics, the non-separable operator has coupled the state at $x$ and $y$. 

We can understand what this coupling entails. If we first alter the state at $y$ with another local excitation, we will affect $\braket{\mathcal{O}(z)}$ only when the next excitation is non-separable. Create this first excitation with a local unitary operator $\mathcal{U}(y')$ where $y'$ and $y$ are timelike-separated but $y'$ is spacelike-separated from $x$ and $z$. Point $y'$ is earlier in time than $y$. The operator $\mathcal{U}(y')$ when acted alone changes the state at $y$, but not $x$ or $z$. The expectation value $\braket{\mathcal{O}(z)}$ does not change for the separable excitation:
\begin{align*}
\braket{\Psi| \mathcal{U}^\dagger (y') \mathcal{U}_s^\dagger(x,y)\mathcal{O}(z) \mathcal{U}_s(x,y)\mathcal{U}(y')|\Psi}
&=
\braket{\Psi|e^{i\mathcal{O}(x)}\mathcal{O}(z) e^{-i\mathcal{O}(x)} e^{i\mathcal{O}(y')} e^{i\mathcal{O}(y)} e^{-i\mathcal{O}(y)} e^{-i\mathcal{O}(y')}|\Psi}
\\
&=
\braket{\Psi|e^{i\mathcal{O}(x)}\mathcal{O}(z) e^{-i\mathcal{O}(x)} |\Psi}.
\numberthis
\end{align*}
The expectation value $\braket{\mathcal{O}(z)}$ does change for the non-separable excitation:
\begin{align*}
\braket{\Psi| \mathcal{U}^\dagger (y') \mathcal{U}_{ns}^\dagger(x,y)\mathcal{O}(z) \mathcal{U}_{ns}(x,y)\mathcal{U}(y')|\Psi}
&=
\braket{\Psi|\mathcal{U}^\dagger (y') (\mathcal{O}(z) + i\mathcal{O}(y) [\mathcal{O}(x) , \mathcal{O}(z)] +\ldots ) \mathcal{U}(y')|\Psi}
\\
&=
\braket{\Psi|\mathcal{O}(z) + i\mathcal{U}^\dagger (y') \mathcal{O}(y)\mathcal{U}(y')  [\mathcal{O}(x) , \mathcal{O}(z)] +\ldots |\Psi}
\end{align*}
As $y$ and $y'$ are timelike-separated, the above expectation value of an operator at $z$ has changed in response to an excitation at $y'$ which is spacelike-separated from $z$. 

The behavior we have identified for non-separable excitations violates the causality properties of local quantum field theory at the time the operator acts. Measurements at $z$ are affected by a local excitation at $y'$, which was spacelike-separated from $z$. This violation of causality is no surprise, as acting with $\mathcal{U}_{ns}$ corresponds to turning on an interaction $\mathcal{O}(x) \mathcal{O}(y)$ in the Hamiltonian, which couples the field at spacelike-separated points. This term is not allowed in the Hamiltonian of a local quantum field theory. Even a non-local experimentalist in the closed system cannot act with this operator as long as the theory that describes the experimentalist and system are both local quantum field theories. We conclude that there is a restriction on localized operators and excitations in local quantum field theories: the operators and excitations must be separable. Separability can be tested using the criteria we have used in this section.

Separable unitary operators localized to two different regions of a Cauchy surface do not change the entanglement entropy of those regions, as the operators can be expressed as the product of unitary operators, each local in a different subregion. For example, $\mathcal{U}_s(x,y)$ does not change the entanglement entropy of region $A$ or $B$ for $x \in A, y \in B$. 

\subsection{Noteworthy quantities as localized unitary operators}

We give examples of familiar quantities that are localized unitary operators. Squeezed states, coherent states, generalized coherent states are examples of excitations that can be created by localized unitary operators, and their interpretations are well-understood \cite{Walls83, Ueda93, Gilmore72,Perelomov72}. In a free $(d+1)$-dimensional field theory, a coherent state is (cf. \cite{Yaffe82})
\begin{equation}
\ket{\Psi_c(\pi_c(\mathbf{x}),\phi_c(\mathbf{x}))} = e^{i\int d^d x \pi_c(\mathbf{x}) \hat{\phi}(\mathbf{x}) - \phi_c(\mathbf{x}) \hat{\pi}(\mathbf{x})}\ket{0}.
\end{equation}
The state is labelled by its expectation values 
\begin{align}
\braket{\Psi_c(\pi_c(\mathbf{x}),\phi_c(\mathbf{x}))|\phi(\mathbf{y})|\Psi_c(\pi_c(\mathbf{x}),\phi_c(\mathbf{x}))} &= \phi_c(\mathbf{y}).
\\
\braket{\Psi_c(\pi_c(\mathbf{x}),\phi_c(\mathbf{x}))|\pi(\mathbf{y})|\Psi_c(\pi_c(\mathbf{x}),\phi_c(\mathbf{x}))} &= \pi_c(\mathbf{y}).
\end{align}
A coherent state is a localized excitation when both $\phi_c(\mathbf{x}),\pi_c(\mathbf{x})$ have compact support. Generalized coherent states are analogous constructions for arbitrary Lie groups \cite{Perelomov77} and also create localized excitations.

Path-ordered exponentials can be localized unitary operators for certain choices of path. If the path is nowhere spacelike, then the path ordering is a time ordering. If the path is everywhere spacelike, then the operator creates an excitation at a single time. Wilson loops with spacelike integration paths are examples of these operators, and they create flux tubes along their path.

Localized unitary operators model certain quantum quenches exactly. Quantum quenches are abrupt changes in the Hamiltonian. For instance, the coefficient $\lambda(x)$ of some operator $\mathcal{O}(x)$ in the Hamiltonian may change suddenly. If the state was the ground state of the Hamiltonian, the state after the quench is an excited state of the new Hamiltonian. In global quenches, $\lambda(x)$ is constant in space. In inhomogeneous quenches, $\lambda(x)$ varies in space. Two different types of quenches go by the name ``local quenches''. One type of local quench involves preparing two different states in half the space, joining them, and evolving with time \cite{EislerP07,CalabreseC207}. Another type of local quench is given by changing $\lambda(x)$ at one spacetime point \cite{NozakiNT14,HeNTW14,Nozaki14}. We give the localized unitary operators that describe this second type of local quench. Global and inhomogeneous quenches are described in a similar way, although the corresponding unitary operators are fully non-localized. In a $(d+1)$-dimensional field theory,
\begin{align}
\text{Global quench}:~ & \mathcal{U} = e^{-i \int d^d x \mathcal{O}(x)}.
\\
\text{Inhomogeneous quench}:~ & \mathcal{U} = e^{-i\int d^d x f(x) \mathcal{O}(x)}.
\\
\text{Local quench}:~ & \mathcal{U} = e^{-i\mathcal{O}(x)}.
\end{align}
Motivated by the properties of non-separable unitary operators, we see that non-separable operators model a ``non-separable quantum quench''
\begin{equation}
\text{Non-separable quench}: ~ \mathcal{U} = e^{-i \mathcal{O}(x) \mathcal{O}(y)}.
\end{equation}
Introducing a small parameter $\alpha$ in the exponent to control the strength of the quench, the first, second, and third-order corrections to operator expectation values are straightforward to calculate in CFTs as they often involve two, three, and four-point functions.

\subsection{Localized non-unitary operators do not always create localized excitations}

In this section, we show that the familiar local operators in field theory do not always create localized excitations. While this may seem counter-intuitive, locality in field theory is enforced through commutators, and expectation values in states $\mathcal{O}\ket{\Psi}$ do not involve any commutators with $\mathcal{O}$. Moreover, the statement that two operators commute at spacelike separation is not a statement about expectation values in the state created by acting with one of those operators. In section 5, we showed that localized finite-norm operators generically create non-localized excitations. An operator $\mathcal{O}$ has a finite norm if $\mathcal{O}\ket{\Psi}$ has a finite norm for all normalized states $\ket{\Psi}$. A related conclusion is that local non-unitary operators do not always model a local quench exactly. Infinite-norm operators will be treated more carefully, and we give a specific infinite-norm operator that creates a non-local excitation.

We first address an intuition that is sometimes held about local operators creating local excitations. Consider a real scalar field in $(3+1)$ dimensions. We may ask about the interpretation of the state
\begin{equation}
\phi(\mathbf{x})\ket{0} = \int \frac{d^3p}{(2\pi)^3} \frac{1}{2 E_\mathbf{p}} e^{-i \mathbf{p} \cdot \mathbf{x}} \ket{\mathbf{p}}.
\end{equation}
We will paraphrase the interpretation of this state given in a well-known field theory textbook \cite{Peskin}. For small (non-relativistic) $\mathbf{p}$, $E_\mathbf{p}$ is approximately constant, and in this case $\phi(\mathbf{x})\ket{0}$ approaches the non-relativistic expression for a position eigenstate $\ket{\mathbf{x}}$ in basis $\ket{\mathbf{p}}$. To quote the authors, ``we will therefore put forward the same interpretation, and claim that the operator $\phi(\mathbf{x})$, acting on the vacuum, creates a particle at position $\mathbf{x}$.'' Moreover, this interpretation is corroborated by calculating
\begin{align}
\braket{0|\phi(\mathbf{x})|\mathbf{p}} 
&= \braket{0|
\int \frac{d^3p'}{(2\pi)^3} 
\frac{1}{\sqrt{2 E_{\mathbf{p}'}}}
\left(
a_{\mathbf{p}'}
e^{i \mathbf{p}' \cdot \mathbf{x}}
+
a_{\mathbf{p}'}^\dagger
e^{-i \mathbf{p}' \cdot \mathbf{x}}
\right)
\sqrt{2 E_\mathbf{p}} a_\mathbf{p}^\dagger|0}
\\
&=e^{i \mathbf{p} \cdot \mathbf{x}}.
\end{align}
This is the same as the inner product $\braket{\mathbf{x}|\mathbf{p}}$ in non-relativistic quantum mechanics. We may perform another check to learn that the analogy with quantum mechanics is only valid in the non-relativistic limit.
\begin{align}
\text{QM}: &\quad \braket{\mathbf{x}|\mathbf{y}} = \delta^{(3)}(\mathbf{x} - \mathbf{y})
\\
\text{QFT}:&\quad  \braket{0|\phi(\mathbf{x}) \phi(\mathbf{y})|0} = \int \frac{d^3p}{(2\pi)^3} 
\frac{e^{i  \mathbf{p} \cdot (\mathbf{x}-\mathbf{y}) }}{2E_{\mathbf{p}}}      
=D(\mathbf{x}-\mathbf{y}) .
\end{align}
In the non-relativistic approximation, $E_\mathbf{p}$ is approximately constant, and both expressions are delta functions. The authors of course never make an erroneous claim, for example that $\phi(\mathbf{x})$ creates a particle only at $\mathbf{x}$ but nowhere else. In fact, particles themselves are approximate notions, and it has been shown that localizing a finite number of particles in a single region is in tension with causality \cite{KnightLocalization,NewtonWigner,Hegerfeldt74}. We have reproduced a textbook argument here to make explicit what considerations and terminology may lead one to the incorrect intuition that if a field theory operator is local it creates a local excitation.

In field theory, the Dirac orthogonality condition does not diagnose locality as we have defined it. The condition that the inner product between two states $\braket{0|\phi(x) \phi(y)|0} = D(x-y)$ grows small as the separation between $x,y$ grows large is known as asymptotic locality \cite{AsymptoticLocality}. We have seen how localized unitary operators create localized excitations, and even localized unitary operators are not Dirac orthogonal. Consider operators of the form $\mathcal{U} = e^{-i \mathcal{O}}$:
\begin{equation}
\braket{\Psi|\mathcal{U}^\dagger(y) \mathcal{U}(x)|\Psi}  = \braket{\Psi|\Psi} +i\braket{\Psi|\mathcal{O}(y) - \mathcal{O}(x)|\Psi}+\ldots
\end{equation}
The failure of local excitations in field theory to obey the Dirac orthogonality condition illustrates that not all quantum-mechanical measures of locality are useful measures in field theory.

Our discussion of the Reeh-Schlieder theorem in section 5.3 applies to field theory as well. The Reeh-Schlieder theorem is a consequence of local non-unitary operators acting in entangled states. It is widely accepted that the source of Reeh-Schlieder theorem in field theory is the entanglement between spatial regions \cite{Redhead95}. To make the connection with our quantum-mechanical explanation of the Reeh-Schlieder theorem, we should consider finite-norm operators in field theory localized to some region. Local operators generically have infinite norm, and must be smeared over some region to have finite norm. Just as in quantum mechanics, these finite-norm non-unitary operators may be localized, but they create the non-localized excitations described by the Reeh-Schlieder theorem.

Some local infinite-norm operators create non-localized excitations. For example, consider the infinite-norm operator
\begin{equation}
\mathcal{O}_e = e^{\alpha\mathcal{O}(x)}.
\end{equation}
Here, $\alpha$ is real and $\mathcal{O}$ is Hermitian. Consider how the expectation value of some operator in the vacuum $\braket{0|\mathcal{O}'|0}$ changes in the excited state $\mathcal{O}_e\ket{0}$. To first order in $\alpha$ this excitation does not change the state's norm as $\braket{0|\mathcal{O}(x)|0} = 0$. So, for the first-order calculation, we do not have to regulate the operator. The expectation value to first order is therefore
\begin{equation}
\braket{0|\mathcal{O}_e(x) \mathcal{O}'(y) \mathcal{O}_e(x) |0} \approx 
\braket{0|\mathcal{O}'|0} 
+
\alpha \braket{0|\left\{\mathcal{O}(x), \mathcal{O}'(y)\right\}|0}
+\ldots
\end{equation}
The anticommutator of two operators does not vanish for spacelike separations and so this local infinite-norm operator creates a non-localized excitation.

\subsection{Certain local non-unitary operators create local excitations}

While not all infinite-norm local operators create local excitations, we show some which do. The locality of these operators comes from the singularity structure of the infinite-norm states they create. This is in contrast to unitary operators, which obtain their locality through operator commutators. To calculate correlators in excitations created by infinite-norm operators, we must first regulate the norm. One way to regulate is to dampen the high-energy modes, which is equivalent to inserting the operator at complex time \cite{CalabreseC05,NozakiNT14}: 
\begin{equation}
e^{-\delta H} \mathcal{O}(x) \ket{0} = \mathcal{O}(x-i\delta)\ket{0}.
\end{equation}
We have used the shorthand $x \pm i\delta \equiv (t \pm i\delta, \mathbf{x})$. Expectation values are taken with the limit $\delta \rightarrow 0$.

As a simple example of an expectation value in an infinite-norm state, we consider the two-point function $\braket{\phi(x) \phi(y)}$ of a free scalar field in state $\phi(z)\ket{0}$, and we work in $d+1>2$ spacetime dimensions. We will find that $\phi(z)$ creates a local excitation at $z$. Suppose $x,y$ are spacelike-separated from $z$. This way the $\delta \rightarrow 0$ limit can be taken without crossing any branch cuts in complex time, and so including the state normalization factor,
\begin{align*}
\braket{\phi(x) \phi(y)}
&\equiv \frac{
\braket{
0|
\phi(z+i\delta) \phi(x) \phi(y) \phi(z-i\delta)
|0
}
}
{
\braket
{0|
\phi(z+i\delta)\phi(z-i\delta)
|0}
}
\\
&=
\frac{D(z+i\delta,x) D(y,z-i\delta) + D(z+i\delta,y)D(x,z-i\delta)}{D(z+i\delta,z-i\delta)}
+
D(x,y).
\numberthis
\end{align*}
Here, $D(x-y) = \braket{0|\phi(x)\phi(y)|0}$. The two-point function is unchanged by the excitation in comparison to its vacuum expectation value, as only the $\phi(z+i\delta) \phi(z-i\delta)$ contraction in the numerator has the same divergence as the denominator in the $\delta \rightarrow 0$ limit, as long as $x,y$ are spacelike-separated from $z$:
\begin{equation}
\lim_{\delta \rightarrow 0} \braket{\phi(x) \phi(y)}= D(x,y).
\end{equation}
The same conclusion holds for an n-point function in this state. The infinite-norm local operator $\phi(z)$ creates a local excitation at $z$ even though it is not unitary.

In general CFTs, we can prove the same behavior we saw in the free scalar case, that a non-unitary infinite-norm local operator with definite conformal dimension creates a local excitation. When calculating a correlation function in the state $\mathcal{O}(x-i\delta)\ket{\Psi}$, the OPE between $\mathcal{O}$ and $\mathcal{O}^\dagger$ may be used when $x$ is spacelike separated from the locations of the other operators in the correlation function. When $\ket{\Psi}$ is a conformally-invariant state, for example the vacuum, the identity dominates the OPE in the $\delta \rightarrow 0$ limit, and the correlator is unaffected by the excitation created by $\mathcal{O}$. We will see an explicit example of this process in section 7.2. This argument applies when $\mathcal{O}$ has definite non-zero conformal dimension. For example, this excludes the operator $\phi$ of the free scalar in $(1+1)$ dimensions, which has conformal dimension zero and creates an excitation that is not asymptotically local \cite{AsymptoticLocality}. Explicitly, for $x$ spacelike-separated from all $y_i$,
\begin{align*}
\braket{ \mathcal{O}(y_1) \mathcal{O}(y_2) \ldots \mathcal{O}(y_n) }
&\equiv
\frac{
\braket{\Psi|\mathcal{O}^\dagger (x+i\delta) \mathcal{O}(y_1) \mathcal{O}(y_2) \ldots \mathcal{O}(y_n) \mathcal{O}(x-i\delta) |\Psi}
}
{
\braket{\Psi|\mathcal{O}^\dagger (x+i\delta) \mathcal{O}(x-i\delta)|\Psi}
}
\\
&=
\sum_{\Delta_k, s_k}
\frac{
\braket{ \Psi| \left[ 
C_{\mathcal{O}^\dagger \mathcal{O} \mathcal{O}_k }
 \frac{\mathcal{O}_k(x+i\delta)}{(2i\delta)^{2\Delta-\Delta_k}}
\right]
\mathcal{O}(y_1) \mathcal{O}(y_2) \ldots \mathcal{O}(y_n)
|\Psi}
}
{
\braket{\Psi|\mathcal{O}^\dagger (x+i\delta) \mathcal{O}(x-i\delta)|\Psi}
}.
\numberthis
\label{CFTOPEDivergence}
\end{align*}
The sum is over all operators $\mathcal{O}_k$, which we have indexed by dimension $\Delta_k$ and spin $s_k$, and the $C_{\mathcal{O}^\dagger \mathcal{O} \mathcal{O}_k }$ are theory-dependent coefficients. If $\mathcal{O}_k$ is a primary operator, then $C_{\mathcal{O}^\dagger \mathcal{O} \mathcal{O}_k }$ is the three-point coefficient.

As the state $\ket{\Psi}$ is conformally invariant, $\braket{\Psi|\mathcal{O}_p|\Psi} = 0$ for all local primary operators $\mathcal{O}_p$. Therefore, $\mathcal{O}^\dagger \mathcal{O}$ must contain the identity in its OPE in order for the two-point function in this state, $\braket{\Psi|\mathcal{O}^\dagger (x+i\delta) \mathcal{O}(x-i\delta)|\Psi}$, to be non-zero. We will assume the identity is present in this OPE. It follows that only the identity's contribution to the OPE in the numerator of \eqref{CFTOPEDivergence} survives the $\delta \rightarrow 0$ limit.
\begin{align*}
\braket{ \mathcal{O}(y_1) \mathcal{O}(y_2) \ldots \mathcal{O}(y_n) }
&=\sum_{\Delta_k, s_k}
\frac{
\braket{ \Psi|\left[ 
C_{\mathcal{O}^\dagger \mathcal{O} \mathcal{O}_k }
 \frac{\mathcal{O}_k(x+i\delta)}{(2i\delta)^{2\Delta-\Delta_k}}
\right]
\mathcal{O}(y_1) \mathcal{O}(y_2) \ldots \mathcal{O}(y_n)
|\Psi}
}
{
\braket{\Psi|\mathcal{O}^\dagger (x+i\delta) \mathcal{O}(x-i\delta)|\Psi}
}
\\
&
=
\sum_{\Delta_k, s_k}
\frac{
\braket{ \Psi|\left[ 
C_{\mathcal{O}^\dagger \mathcal{O} \mathcal{O}_k }
 \frac{\mathcal{O}_k(x+i\delta)}{(2i\delta)^{2\Delta-\Delta_k}}
\right]
\mathcal{O}(y_1) \mathcal{O}(y_2) \ldots \mathcal{O}(y_n)
|\Psi}
}
{
(2i\delta)^{-2\Delta}
}
\\
&\stackrel{\delta \rightarrow 0}{\longrightarrow}
\braket{\Psi|\mathcal{O}(y_1) \mathcal{O}(y_2) \ldots \mathcal{O}(y_n)|\Psi}
.
\numberthis
\label{CFTLocalityArgument}
\end{align*}

Even if the field theory is not a CFT, an argument similar to \eqref{CFTLocalityArgument} shows that $\mathcal{O}(x)$ creates local excitations if we assume a certain short-distance factorization. The only non-zero contribution to a correlator evaluated a state created by $\mathcal{O}(x-i\delta)$ comes from the $\delta \rightarrow 0$ contribution. If $x$ is spacelike from the insertion points of all the other operators, then the $\delta \rightarrow 0$ limit does not cross any branch cut of the complex-time correlator. If, in a particular state of a field theory, the correlator factorizes on the $\delta \rightarrow 0$ singularity, then $\mathcal{O}(x)$ creates a local excitation, and the argument proceeds similar to the CFT case:
\begin{align*}
\frac{
\braket{\mathcal{O}^\dagger (x+i\delta) \mathcal{O}(y_1) \mathcal{O}(y_2) \ldots \mathcal{O}(y_n) \mathcal{O}(x-i\delta)}
}
{
\braket{\mathcal{O}^\dagger (x+i\delta) \mathcal{O}(x-i\delta)}
}
&\stackrel{\delta \rightarrow 0}{\longrightarrow}\frac{\braket{\mathcal{O}^\dagger (x+i\delta) \mathcal{O}(x-i\delta)}
\braket{\mathcal{O}(y_1) \mathcal{O}(y_2) \ldots \mathcal{O}(y_n)}
}
{
\braket{\mathcal{O}^\dagger (x+i\delta) \mathcal{O}(x-i\delta)}
}
\\
&=
\braket{\mathcal{O}(y_1) \mathcal{O}(y_2) \ldots \mathcal{O}(y_n)}, 
~~ y_i-x ~ \text{spacelike}.
\numberthis
\end{align*}

Strictly speaking, local operators themselves are only operator-valued distributions. While certain infinite-norm operators may create local excitations, these operators must be smeared with some test function to create a physical excitation with finite norm. But once the non-unitary operator has finite norm, the finite-norm excitation will generically not be localized to the region of smearing. As such, the conclusions drawn from the locality properties of infinite-norm operators must be treated with care, as they may not extend to the operators' smeared counterparts.

\section{Entanglement Entropy in Excited States}

Entanglement entropy has recently emerged as a useful probe of excited-state dynamics in $(1+1)$-dimensional conformal field theories. A compelling quasi-particle picture has been proposed for local operators, wherein generic local operators create local excitations that can be interpreted as entangled pairs of quasi-particles \cite{NozakiNT14}. In this section, we revisit the results in the literature and show how, while some infinite-norm operators create local excitations that may admit a quasi-particle description, entanglement and Renyi entropies change non-locally for other infinite-norm operators, and so the quasi-particle picture is not universal for all local operators. It is known that the quasi-particle picture fails for some theories \cite{Leichenauer15,AsplundBGH115}, and we show its failure in theories in which the picture is expected to be accurate. We show the results in the literature are consistent with evidence we have presented that local operators with definite conformal dimensions create local excitations. We also prove a causality relation for entanglement entropy.

\subsection{Causal properties of entanglement entropy}

The details of how localized excitations are implemented by localized unitary operators motivate a general causality condition for entanglement entropy in quantum field theories. The condition applies to pure states. Our result extends the result proved in ref. \cite{Headrick14}. This earlier result makes use of the fact that, for a localized excitation within domain of dependence $\mathcal{D}(A)$ of subregion $A$ of a Cauchy surface, one can always find a Cauchy surface $A'$ of $\mathcal{D}(A)$ such that the state on $A'$ is unaffected by the excitation. The excitation's support $R$ is in the future of $A'$ and to the past of $A$. As a reminder, the domain of dependence is defined as the region $\mathcal{D}(A)$ that, if an inextendible curve that is nowhere spacelike passes through the region, then this curve must intersect $A$. The domain of dependence $\mathcal{D}(A') = \mathcal{D}(A)$. The reduced density matrices on $A, A'$ are unitarily related and so the entanglement entropy does not change.

However, having a Cauchy surface $A'$ that is unaffected by the excitation is not a necessary condition. For example, no such $A'$ exists for local excitations prepared by ingoing excitations, yet these excitations still do not change entanglement entropy. We provide a proof that does not rely on the existence of $A'$, but on the properties of the excitation regardless of how it was prepared.

Consider a quantum field theory in some pure state $\ket{\Psi}$. We choose a purely spatial surface at time $t$ as a Cauchy surface for simplicity. Divide the spatial surface into two regions $A, B$ with reduced density matrices $\rho_A, \rho_B$. Consider an excitation localized within $A$ at time $t$. This excitation can be created by acting with a unitary operator $\mathcal{U}(A)$ localized in A. This localized excitation does not change $\rho_B$. The entanglement entropy $S_B$ of region $B$ therefore does not change either. As the state $\ket{\Psi}$ is pure, $S_A = S_B$, and so $S_A$ does not change. If the perturbation is localized within $B$, the excitation does not change $\rho_A$ or $S_A$.

Only when the excitation is localized to a region that is causally connected to both $A$ and $B$ will these arguments fail. In this case, the entanglement entropy may change. The complement of $\mathcal{D}(A) \cup \mathcal{D}(B)$ to the past of $t$ is precisely the correct region. This region includes its boundary, which consists of null rays. We recover the causality condition of ref. \cite{Headrick14}. Some excitations with support in both $A$ and $B$ can leave the entanglement entropy unchanged. As we showed in section 6.2, the excitations created by separable unitary operators accomplish this.

\subsection{Entanglement entropy calculations with infinite-norm operators}

In light of our conclusions that some infinite-norm local operators can create non-localized excitations, the results of recent entanglement entropy calculations may seem surprising. We show how these calculations are consistent with our results.

Calculations of entanglement entropy in excited states created by infinite-norm operators have shown that Renyi and entanglement entropies change only when the operator insertion is null or timelike to the subregion \cite{HeNTW14, Nozaki14,Caputa14,NozakiNT14,AsplundBGH115,AsplundBGH215,CaputaV15,CaputaSST14}. States of the form $\mathcal{O}(x) \ket{0}$ were considered in $(1+1)$-dimensional CFTs. It was suggested that the jumps in entanglement entropy reveals a local quasi-particle picture. In this picture, a local operator creates quasi-particle pairs that propagate at the speed of light from the operator's insertion point. Entanglement entropy changes only when one member of the pair is inside the subregion, but not both members.

The $(1+1)$-dimensional calculations we address use the replica trick to calculate entanglement entropy. In the replica trick, entanglement entropy of interval $A$ is calculated from the replicated density matrix $\text{tr}\rho^n_A$, and conveniently given by correlators with twist operators $\Phi_n$ \cite{ReplicaTrick,Holzhey94}. The path integral for a field $\phi$ on an $n$-sheeted Riemann surface is given by a path integral for fields $\phi_i$ living on $\mathbb{C}$ with certain boundary conditions relating the $\phi_i$. These boundary conditions can be represented by inserting twist operators at the endpoints of the interval. Twist operators are primary. Correlators are taken in the theory with the $n$ fields $\phi_i$. For details, see ref. \cite{CalabreseC09}.

We consider a single interval $A$ with endpoints $u,v$. The replica trick for excited states has been established \cite{HeNTW14, Nozaki14,Caputa14,NozakiNT14}. Take $\mathcal{O}$ to be an operator creating an excited state. For example, one such operator could be $\mathcal{O}(x) = \prod_i^n \phi_i(x)$. Renyi entropies are calculated from 
\begin{equation}
\text{Tr}(\rho_A^n) =\frac{\braket{0| \mathcal{O}^\dagger(x+i \delta)\Phi_n(u)\bar{\Phi}_n(v)\mathcal{O}(x-i\delta)|0}
}
{
\braket{ 0| \mathcal{O}^\dagger(x+i\delta)\mathcal{O}(x-i\delta)|0}
}.
\end{equation}
The normalization is such that $\text{Tr}(\rho_A^n) = 1$ for $n = 1$. Entanglement entropy is calculated from the Renyi entropy. In generic excited states, when the Renyi entropy changes, the entanglement entropy will change as well.

Suppose $\mathcal{O}$ is an operator with definite conformal dimension. If $x$ is spacelike-separated from $u,v$, we can use the $\mathcal{O}(x+i\delta) \mathcal{O}^\dagger(x-i\delta)$ OPE to understand what happens in the limit $\delta \rightarrow 0$. For finite $\delta$ the excitation created by $\mathcal{O}$ has a finite-norm and can be non-local. Indeed, entanglement entropy changes at spacelike separations for finite $\delta$ \cite{HeNTW14, Nozaki14,Caputa14,NozakiNT14}. The leading contribution to the OPE for small $\delta$ is from the identity operator, and we showed in \eqref{CFTLocalityArgument} how this implies the locality of certain operator excitations. We will revisit and provide context for this statement shortly, comparing it to the result in ref. \cite{AsplundBGH215} to understand when the leading contribution to the full correlator comes from the identity and when it can come from the full identity block. For small $\delta$,
\begin{equation}
\text{Tr}(\rho_A^n) = \frac{\braket{0| \mathcal{O}^\dagger(x+i \delta)\mathcal{O}(x-i\delta)|0} \braket{0|\Phi_n(u)\bar{\Phi}_n(v)|0} + \text{subleading}
}
{
\braket{ 0| \mathcal{O}^\dagger(x+i\delta)\mathcal{O}(x-i\delta)|0}
}.
\end{equation}
For $\delta \rightarrow 0$,
\begin{equation}
\text{Tr}(\rho_A^n) = \braket{0|\Phi_n(u)\bar{\Phi}_n(v)|0}.
\label{RenyiUnchanged}
\end{equation}
The excitation created by $\mathcal{O}$ does not affect the Renyi entropy. This argument was also given in section 6.5.

As $x$ becomes null-separated from $u$ or $v$, the OPE of $\mathcal{O}^\dagger(x+i\delta)\mathcal{O}(x-i\delta)$ is not convergent because the twist operators are within what would be the neighborhood of convergence. In ref. \cite{AsplundBGH215}, the authors instead consider the vacuum block approximation to the four-point function, which is valid under certain assumptions and in a particular limit. They observe that this function has a certain branch cut, that when performing the continuation to real time, causes the entanglement entropy to pick up an additional contribution when the excitation is not spacelike-separated from the subregion. This is an example of how the entanglement entropy changes when the subregion becomes null and timelike to $x$. 

Our statement that for spacelike-separations the identity operator and not also its descendants dominates the correlation function as $\delta \rightarrow 0$ is consistent with the recent calculations in ref. \cite{AsplundBGH215} of Renyi and entanglement entropies in the presence of a local operator excitation of the vacuum. In the $\delta \rightarrow 0$ limit, their expression for $\text{Tr}(\rho_A^n)$ reduces to the two-point function of twist operators in the vacuum as long as the excitation is spacelike-separated from the interval. The Renyi entropy is therefore unchanged by the excitation, just as we found in \eqref{RenyiUnchanged}. In the expression for the vacuum conformal block used in ref. \cite{AsplundBGH215}, $\delta \rightarrow 0$ is the $z \rightarrow 1$ limit. The leading divergence in the vacuum block corresponds to exchanging the identity, while all divergences subleading in $z-1$ correspond to exchanging descendants of the identity. As $\delta \rightarrow 0$, only the leading divergence to the vacuum block gives a non-zero contribution to the correlator. Only when the excitation is not spacelike-separated from the interval does $\bar{z}$ pass to its second sheet, and the branch cut in the conformal block causes the Renyi entropy to change.

The argument we have given that infinite-norm local operators create local excitations fails when $\mathcal{O}$ does not have a definite scaling dimension. As an example, instead excite the vacuum with the operator 
\begin{equation}
\mathcal{O}_e = \prod_i^n e^{\alpha \mathcal{O}_i(z)}
\end{equation}
Here $\alpha$ is real and contains a small dimensionless parameter. For simplicity, take $\mathcal{O}_i = \mathcal{O}^\dagger_i$. To first order in $\alpha$, 
\begin{equation}
\mathcal{O}_e = 1+\alpha \sum_i^n \mathcal{O}_n(z) \equiv 1+\alpha \mathcal{O}(z).
\end{equation} 
We denote $\sum_i^n \mathcal{O}_n(z) = \mathcal{O}(z)$ for short. Notice that this operator does not change the state's norm to first order in $\alpha$. The correction to the Renyi entropy is proportional to $\braket{0|\left\{ \mathcal{O}(z), \Phi_n(u) \bar{\Phi}_n(v) \right\}|0}$, and unless the three-point function vanishes, the anti-commutator generically is not zero for $z$ spacelike-separated from $u,v$. For an explicit example, choose $\mathcal{O} = \sum_{i}^n T_n(z)$, the stress tensor. The anticommutator is known \cite{ReplicaTrick}. This calculation may be performed with the entanglement first law. Alternatively, replace $\mathcal{O}_i(z)$ with a non-local operator $\mathcal{O}_i(z_1) \mathcal{O}_i(z_2)$ to see a case in which the Renyi entropy will be non-zero. 

The argument we gave that uses the OPE to show that Renyi and entanglement entropies change in response to a local excitation does not apply to $\mathcal{O}_e$. For example, to first order in $\alpha$, the four-point function is a three-point function involving one $\mathcal{O}$, and so there is no $\mathcal{O} \mathcal{O}$ OPE to take. Said another way, as the OPE $\mathcal{O}_e(z+i\delta) \mathcal{O}_e(z-i\delta)$ contains no divergence to first order in $\alpha$, the contribution of the identity operator to the OPE does not determine the correlator's behavior. While we must introduce a regulator $\delta$ for the state's infinite norm, we need not introduce $\delta$ if we are working to first order in $\alpha$.

The calculations we have shown are consistent with our arguments in section 6.5, as operators $\mathcal{O}$ which have definite conformal dimension change entanglement entropy only when $\mathcal{O}$ is in causal contact with the interval.

We have shown the quasi-particle picture does not describe excitations created by all local operators, but we have provided evidence that operators with definite conformal dimension have a quasi-particle interpretation for certain conformal field theories. Others have demonstrated that the quasi-particle picture is invalid for some field theories \cite{Leichenauer15,AsplundBGH115}. The quasi-particle picture remains a compelling description of certain excitations in certain theories, and understanding its origin may reveal important properties of entanglement entropy.

\section{Discussion}

In this work, we have shown how localized unitary operators are fundamental building blocks of time-dependent quantum systems in entangled states. Localized unitary operators create localized excitations. We have detailed various features of localized unitary operators, including their locality properties, their behavior under superposition, and the difference between separable and non-separable unitary operators. We found that non-separable unitary operators, and their associated non-separable localized excitations, are in conflict with the principles of local quantum field theory. We gave a criterion to test for separability.

We have shown how, unlike local unitary operators, local non-unitary operators can create non-local excitations in entangled states. As a reminder, generic states in field theory are entangled over spatial regions. Local non-unitary operators are state-dependent and can have infinite norm. We provided an example of an infinite-norm local non-unitary operator that creates a non-local excitation. We gave arguments that suggest that certain infinite-norm local non-unitary operators do create local excitations. However, these operators must be smeared to have finite norm, and the resulting finite-norm operators can create fully non-localized excitations. Consequently, one must be careful when drawing conclusions about locality properties based on those of infinite-norm local operators. In practice, however, correlators in excited states created by a non-unitary operator $\mathcal{O}(x)$ can be simpler to calculate than correlators in excited states created by a unitary operator $e^{i\alpha\mathcal{O}(x)}$, which can involve perturbation theory in $\alpha$ and a treatment of divergences.

We defined a local picture for quantum systems that unifies several different manifestations of locality and causality into a simple description. The local picture follows naturally from real-time perturbation theory and the definitions of open and closed quantum systems. According to the local picture, experimentalists can only act through deforming the Hamiltonian, and localized experimentalists can only deform the Hamiltonian locally. Localized unitary operators are central features of the local picture. Deforming the Hamiltonian in a localized region is equivalent to acting with a localized unitary operator on the state, and this operator will create a localized excitation. Generic non-unitary operators create non-localized excitations, so in order to act with these operators on the state, the experimentalist must be fully non-localized herself.

Using the local picture and our analysis of unitary and non-unitary operators, we distilled more formal results in algebraic quantum field theory into elementary statements in quantum mechanics, and demonstrated their underlying mechanisms. We showed how the non-local state preparation described by the Reeh-Schlieder theorem comes from the fact that local non-unitary operators create non-localized excitations in entangled states. The local picture makes clear how the Reeh-Schlieder theorem is intuitive and consistent with causality. Localized experimentalists can only create localized excitations, and so cannot act with the local non-unitary operators that create non-localized excitations.

We applied our results to entanglement entropy in field theory. We used properties of localized excitations to prove a causality condition for entanglement entropy that extends an earlier result \cite{Headrick14}. Our proof applies to separable excitations and states prepared with ingoing excitations. We addressed recent calculations of entanglement entropy in $(1+1)$-dimensional conformal field theories \cite{HeNTW14, Nozaki14,Caputa14,NozakiNT14,AsplundBGH215}, and provided evidence that the locality properties demonstrated by these calculations are only properties of operators with definite conformal dimension. We showed consistency between these calculations and our conclusions about the locality of operator excitations. We provided an example of a local non-unitary operator that changes entanglement non-locally. While the quasi-particle picture is known to fail in certain theories \cite{Leichenauer15,AsplundBGH115}, we concluded that the quasi-particle picture does not describe excitations created by all local operators in theories in which the picture is expected to hold. Understanding whether the picture applies to all localized excitations may provide insights into entanglement entropy.

We connect our results to the AdS/CFT correspondence in the limit in which the bulk is semiclassical. Non-normalizable modes of bulk fields $\phi$ with dual CFT operators $\mathcal{O}$ are turned on at the boundary by acting with the localized unitary operators $\mathcal{T}\left( e^{-i\int d^{d+1}x \phi_0(x) \mathcal{O}(x)}\right)$ in the CFT. An excitation of the CFT on a Cauchy surface $\mathcal{S}$ is associated with a bulk excitation in $\mathcal{Q}_{\mathcal{S}} \cup \mathcal{S}$, where the causal shadow $\mathcal{Q}_{\mathcal{S}}$ is the set of points spacelike-separated from all points in $\mathcal{S}$. This is because the region $\mathcal{Q}_{\mathcal{S}} \cup \mathcal{S}$ is the union of all possible bulk Cauchy surfaces which intersect the boundary at $\mathcal{S}$, and there is generically no preferred way to choose one of these Cauchy surfaces for the bulk theory. Work on operator reconstruction is fully compatible with the fact that local non-unitary operators generically create non-localized excited states. For instance, for every local Hermitian operator $\mathcal{O}(x)$ there is a unitary operator $e^{i\alpha \mathcal{O}(x)}$ which creates a local excitation at $x$. Recent work sheds light on these considerations through a bulk exploration of the Reeh-Schlieder theorem \cite{Morrison14,BanerjeeBPR16}. Recall that unlike unitary operators, non-unitary operators are state-dependent operators. State-dependent operators in AdS/CFT have been explored in detail \cite{PapadodimasR113,PapadodimasR213,PapadodimasR115,PapadodimasR215}.

We expect that our results, along with our elementary treatment of related discussions in diverse branches of the literature will help clarify investigations into locality, causality, entanglement entropy, and the AdS/CFT duality in the future.

\section{Acknowledgements}

We would like to thank our advisor Per Kraus for invaluable guidance, support, and discussions. We also thank River Snively and Eliot Hijano for enjoyable, helpful conversations and comments on the draft. We thank Detlev Buchholz for stimulating correspondence.

\bibliographystyle{ssg}
\bibliography{refs}

\begingroup\raggedright\begin{thebibliography}{10}

\bibitem{HeNTW14}
S.~He, T.~Numasawa, T.~Takayanagi, and K.~Watanabe, ``{Quantum Dimension as
  Entanglement Entropy in 2D CFTs},''
  \href{http://xxx.lanl.gov/abs/1403.0702}{{\tt 1403.0702}}.

\bibitem{Nozaki14}
M.~Nozaki, ``{Notes on Quantum Entanglement of Local Operators},'' {\em JHEP}
  {\bf 10} (2014) 147, \href{http://xxx.lanl.gov/abs/1405.5875}{{\tt
  1405.5875}}.

\bibitem{Caputa14}
P.~Caputa, M.~Nozaki, and T.~Takayanagi, ``{Entanglement of local operators in
  large-N conformal field theories},'' {\em PTEP} {\bf 2014} (2014) 093B06,
  \href{http://xxx.lanl.gov/abs/1405.5946}{{\tt 1405.5946}}.

\bibitem{NozakiNT14}
M.~Nozaki, T.~Numasawa, and T.~Takayanagi, ``{Quantum Entanglement of Local
  Operators in Conformal Field Theories},'' {\em Phys.Rev.Lett.} {\bf 112}
  (2014) 111602, \href{http://xxx.lanl.gov/abs/1401.0539}{{\tt 1401.0539}}.

\bibitem{AsplundBGH115}
C.~T. Asplund, A.~Bernamonti, F.~Galli, and T.~Hartman, ``{Entanglement
  Scrambling in 2d Conformal Field Theory},'' {\em JHEP} {\bf 09} (2015) 110,
  \href{http://xxx.lanl.gov/abs/1506.03772}{{\tt 1506.03772}}.

\bibitem{AsplundBGH215}
C.~T. Asplund, A.~Bernamonti, F.~Galli, and T.~Hartman, ``{Holographic
  Entanglement Entropy from 2d CFT: Heavy States and Local Quenches},'' {\em
  JHEP} {\bf 02} (2015) 171, \href{http://xxx.lanl.gov/abs/1410.1392}{{\tt
  1410.1392}}.

\bibitem{CaputaV15}
P.~Caputa and A.~Veliz-Osorio, ``{Entanglement constant for conformal
  families},'' {\em Phys. Rev.} {\bf D92} (2015), no.~6 065010,
  \href{http://xxx.lanl.gov/abs/1507.00582}{{\tt 1507.00582}}.

\bibitem{CaputaSST14}
P.~Caputa, J.~Simón, A.~Štikonas, and T.~Takayanagi, ``{Quantum Entanglement
  of Localized Excited States at Finite Temperature},'' {\em JHEP} {\bf 01}
  (2015) 102, \href{http://xxx.lanl.gov/abs/1410.2287}{{\tt 1410.2287}}.

\bibitem{InInSchwinger}
J.~Schwinger, ``Brownian Motion of a Quantum Oscillator,'' {\em Journal of
  Mathematical Physics} {\bf 2} (1961), no.~3 407--432.

\bibitem{InInKeldysh}
L.~V. Keldysh, ``{Diagram technique for nonequilibrium processes},'' {\em Zh.
  Eksp. Teor. Fiz.} {\bf 47} (1964) 1515--1527. [Sov. Phys. JETP20,1018(1965)].

\bibitem{ReehSchlieder}
H.~Reeh and S.~Schlieder, ``Bemerkungen zur unit{\"a}r{\"a}quivalenz von
  lorentzinvarianten feldern,'' {\em Il Nuovo Cimento (1955-1965)} {\bf 22}
  (2008), no.~5 1051--1068.

\bibitem{KnightLocalization}
J.~M. Knight, ``Strict Localization in Quantum Field Theory,'' {\em Journal of
  Mathematical Physics} {\bf 2} (1961), no.~4 459--471.

\bibitem{Licht63}
A.~L. Licht, ``Strict Localization,'' {\em Journal of Mathematical Physics}
  {\bf 4} (1963), no.~11 1443--1447.

\bibitem{Licht66}
A.~L. Licht, ``Local States,'' {\em Journal of Mathematical Physics} {\bf 7}
  (1966), no.~9 1656--1669.

\bibitem{Redhead95}
M.~Redhead, ``More ado about nothing,'' {\em Foundations of Physics} {\bf 25}
  (1995), no.~1 123--137.

\bibitem{Hellwig70}
K.~E. Hellwig and K.~Kraus, ``Operations and measurements. II,'' {\em
  Communications in Mathematical Physics} {\bf 16} (1970), no.~2 142--147.

\bibitem{NewtonWigner}
T.~D. Newton and E.~P. Wigner, ``Localized States for Elementary Systems,''
  {\em Rev. Mod. Phys.} {\bf 21} (Jul, 1949) 400--406.

\bibitem{Halvorson00}
H.~Halvorson, ``{Reeh-Schlieder defeats Newton-Wigner: On alternative
  localization schemes in relativistic quantum field theory},'' {\em Phil.
  Sci.} {\bf 68} (2001) 111--133,
  \href{http://xxx.lanl.gov/abs/quant-ph/0007060}{{\tt quant-ph/0007060}}.

\bibitem{Buchholz15}
D.~Buchholz and E.~St{\o}rmer, ``Superposition, Transition Probabilities and
  Primitive Observables in Infinite Quantum Systems,'' {\em Communications in
  Mathematical Physics} {\bf 339} (2015), no.~1 309--325.

\bibitem{Walls83}
D.~F. Walls, ``Squeezed states of light,'' {\em Nature} {\bf 306} (1983) 141.

\bibitem{Ueda93}
M.~Kitagawa and M.~Ueda, ``Squeezed spin states,'' {\em Phys. Rev. A} {\bf 47}
  (Jun, 1993) 5138--5143.

\bibitem{Gilmore72}
R.~Gilmore, ``On the properties of coherent states,'' {\em Rev. Mex. de Fisica}
  {\bf 23} (1972) 143.

\bibitem{Perelomov72}
A.~M. Perelomov, ``Coherent States for Arbitrary Lie Group,'' {\em Commun.
  Math. Phys.} {\bf 26} (1972) 222.

\bibitem{GlauberCoherentState}
R.~J. Glauber, ``Coherent and Incoherent States of the Radiation Field,'' {\em
  Phys. Rev.} {\bf 131} (Sep, 1963) 2766--2788.

\bibitem{HartmanJK15}
T.~Hartman, S.~Jain, and S.~Kundu, ``{Causality Constraints in Conformal Field
  Theory},'' \href{http://xxx.lanl.gov/abs/1509.00014}{{\tt 1509.00014}}.

\bibitem{Leichenauer15}
S.~Leichenauer and M.~Moosa, ``{Entanglement Tsunami in (1+1)-Dimensions},''
  {\em Phys. Rev.} {\bf D92} (2015) 126004,
  \href{http://xxx.lanl.gov/abs/1505.04225}{{\tt 1505.04225}}.

\bibitem{Headrick14}
M.~Headrick, V.~E. Hubeny, A.~Lawrence, and M.~Rangamani, ``{Causality \&
  holographic entanglement entropy},'' {\em JHEP} {\bf 12} (2014) 162,
  \href{http://xxx.lanl.gov/abs/1408.6300}{{\tt 1408.6300}}.

\bibitem{Haag}
R.~Haag, {\em Local quantum physics: fields, particles, algebras}.
\newblock Texts and Monographs in Physics. Springer-Verlag, 1992.

\bibitem{StreaterWightman}
R.~Streater and A.~Wightman, {\em PCT, spin and statistics, and all that}.
\newblock Mathematical physics monograph series. W.A. Benjamin, 1964.

\bibitem{Mathur93}
S.~D. Mathur, ``{Is the Polyakov path integral prescription too
  restrictive?},'' \href{http://xxx.lanl.gov/abs/hep-th/9306090}{{\tt
  hep-th/9306090}}.

\bibitem{Weinberg05}
S.~Weinberg, ``{Quantum contributions to cosmological correlations},'' {\em
  Phys.Rev.} {\bf D72} (2005) 043514,
  \href{http://xxx.lanl.gov/abs/hep-th/0506236}{{\tt hep-th/0506236}}.

\bibitem{Skenderis08}
K.~Skenderis and B.~C. van Rees, ``{Real-time gauge/gravity duality:
  Prescription, Renormalization and Examples},'' {\em JHEP} {\bf 05} (2009)
  085, \href{http://xxx.lanl.gov/abs/0812.2909}{{\tt 0812.2909}}.

\bibitem{Musso06}
M.~Musso, ``{A new diagrammatic representation for correlation functions in the
  in-in formalism},'' {\em JHEP} {\bf 1311} (2013) 184,
  \href{http://xxx.lanl.gov/abs/hep-th/0611258}{{\tt hep-th/0611258}}.

\bibitem{CasiniHR13}
H.~Casini, M.~Huerta, and J.~A. Rosabal, ``{Remarks on entanglement entropy for
  gauge fields},'' {\em Phys. Rev.} {\bf D89} (2014), no.~8 085012,
  \href{http://xxx.lanl.gov/abs/1312.1183}{{\tt 1312.1183}}.

\bibitem{Soni15}
R.~M. Soni and S.~P. Trivedi, ``{Aspects of Entanglement Entropy for Gauge
  Theories},'' {\em JHEP} {\bf 01} (2016) 136,
  \href{http://xxx.lanl.gov/abs/1510.07455}{{\tt 1510.07455}}.

\bibitem{Ma2015}
C.-T. Ma, ``{Entanglement with Centers},'' {\em JHEP} {\bf 01} (2016) 070,
  \href{http://xxx.lanl.gov/abs/1511.02671}{{\tt 1511.02671}}.

\bibitem{Hegerfeldt74}
G.~C. Hegerfeldt, ``Remark on causality and particle localization,'' {\em Phys.
  Rev. D} {\bf 10} (Nov, 1974) 3320--3321.

\bibitem{BhattacharyaHRT14}
J.~Bhattacharya, V.~E. Hubeny, M.~Rangamani, and T.~Takayanagi, ``{Entanglement
  density and gravitational thermodynamics},'' {\em Phys. Rev.} {\bf D91}
  (2015), no.~10 106009, \href{http://xxx.lanl.gov/abs/1412.5472}{{\tt
  1412.5472}}.

\bibitem{Liu13}
H.~Liu and S.~J. Suh, ``{Entanglement Tsunami: Universal Scaling in Holographic
  Thermalization},'' {\em Phys. Rev. Lett.} {\bf 112} (2014) 011601,
  \href{http://xxx.lanl.gov/abs/1305.7244}{{\tt 1305.7244}}.

\bibitem{HubenyRangamaniTakayanagi}
V.~E. Hubeny, M.~Rangamani, and T.~Takayanagi, ``{A Covariant holographic
  entanglement entropy proposal},'' {\em JHEP} {\bf 0707} (2007) 062,
  \href{http://xxx.lanl.gov/abs/0705.0016}{{\tt 0705.0016}}.

\bibitem{RyuTakayanagi}
S.~Ryu and T.~Takayanagi, ``{Holographic derivation of entanglement entropy
  from AdS/CFT},'' {\em Phys.Rev.Lett.} {\bf 96} (2006) 181602,
  \href{http://xxx.lanl.gov/abs/hep-th/0603001}{{\tt hep-th/0603001}}.

\bibitem{Yaffe82}
L.~G. Yaffe, ``Large {N} limits as classical mechanics,'' {\em Rev. Mod. Phys.}
  {\bf 54} (Apr, 1982) 407--435.

\bibitem{Perelomov77}
A.~M. Perelomov, ``Obobshchennye kogerentnye sostoyaniya i nekotorye ikh
  primeneniya,'' {\em Uspekhi Fizicheskikh Nauk} {\bf 123} (1977), no.~9
  23--55.

\bibitem{EislerP07}
V.~Eisler and I.~Peschel, ``Evolution of entanglement after a local quench,''
  {\em Journal of Statistical Mechanics: Theory and Experiment} {\bf 2007}
  (2007), no.~06 P06005.

\bibitem{CalabreseC207}
P.~Calabrese and J.~Cardy, ``Entanglement and correlation functions following a
  local quench: a conformal field theory approach,'' {\em Journal of
  Statistical Mechanics: Theory and Experiment} {\bf 2007} (2007), no.~10
  P10004.

\bibitem{Peskin}
M.~E. Peskin and D.~V. Schroeder, {\em An introduction to quantum field
  theory}.
\newblock Advanced book program. Westview Press Reading (Mass.), Boulder
  (Colo.), 1995.
\newblock Autre tirage : 1997.

\bibitem{AsymptoticLocality}
R.~Haag, ``Quantum Field Theories with Composite Particles and Asymptotic
  Conditions,'' {\em Phys. Rev.} {\bf 112} (Oct, 1958) 669--673.

\bibitem{CalabreseC05}
P.~Calabrese and J.~Cardy, ``{Evolution of entanglement entropy in
  one-dimensional systems},'' {\em J. Stat. Mech.} {\bf 2005} (2005)
  \href{http://xxx.lanl.gov/abs/0503393}{{\tt 0503393}}.

\bibitem{ReplicaTrick}
P.~Calabrese and J.~L. Cardy, ``{Entanglement entropy and quantum field
  theory},'' {\em J.Stat.Mech.} {\bf 0406} (2004) P06002,
  \href{http://xxx.lanl.gov/abs/hep-th/0405152}{{\tt hep-th/0405152}}.

\bibitem{Holzhey94}
C.~Holzhey, F.~Larsen, and F.~Wilczek, ``{Geometric and renormalized entropy in
  conformal field theory},'' {\em Nucl.Phys.} {\bf B424} (1994) 443--467,
  \href{http://xxx.lanl.gov/abs/hep-th/9403108}{{\tt hep-th/9403108}}.

\bibitem{CalabreseC09}
P.~Calabrese and J.~Cardy, ``{Entanglement entropy and conformal field
  theory},'' {\em J.Phys.} {\bf A42} (2009) 504005,
  \href{http://xxx.lanl.gov/abs/0905.4013}{{\tt 0905.4013}}.

\bibitem{Morrison14}
I.~A. Morrison, ``{Boundary-to-bulk maps for AdS causal wedges and the
  Reeh-Schlieder property in holography},'' {\em JHEP} {\bf 05} (2014) 053,
  \href{http://xxx.lanl.gov/abs/1403.3426}{{\tt 1403.3426}}.

\bibitem{BanerjeeBPR16}
S.~Banerjee, J.-W. Bryan, K.~Papadodimas, and S.~Raju, ``{A toy model of black
  hole complementarity},'' \href{http://xxx.lanl.gov/abs/1603.02812}{{\tt
  1603.02812}}.

\bibitem{PapadodimasR113}
K.~Papadodimas and S.~Raju, ``{Black Hole Interior in the Holographic
  Correspondence and the Information Paradox},'' {\em Phys. Rev. Lett.} {\bf
  112} (2014), no.~5 051301, \href{http://xxx.lanl.gov/abs/1310.6334}{{\tt
  1310.6334}}.

\bibitem{PapadodimasR213}
K.~Papadodimas and S.~Raju, ``{State-Dependent Bulk-Boundary Maps and Black
  Hole Complementarity},'' {\em Phys. Rev.} {\bf D89} (2014), no.~8 086010,
  \href{http://xxx.lanl.gov/abs/1310.6335}{{\tt 1310.6335}}.

\bibitem{PapadodimasR115}
K.~Papadodimas and S.~Raju, ``{Local Operators in the Eternal Black Hole},''
  {\em Phys. Rev. Lett.} {\bf 115} (2015), no.~21 211601,
  \href{http://xxx.lanl.gov/abs/1502.06692}{{\tt 1502.06692}}.

\bibitem{PapadodimasR215}
K.~Papadodimas and S.~Raju, ``{Remarks on the necessity and implications of
  state-dependence in the black hole interior},'' {\em Phys. Rev.} {\bf D93}
  (2016), no.~8 084049, \href{http://xxx.lanl.gov/abs/1503.08825}{{\tt
  1503.08825}}.

\end{thebibliography}\endgroup

\end{document}